\documentclass[journal=jcisd8,manuscript=article]{achemso}
\setkeys{acs}{articletitle = true}
\mciteErrorOnUnknownfalse

\usepackage[version=3]{mhchem} 
\usepackage{amsfonts}
\usepackage{makecell}
\usepackage{graphicx}
\usepackage{caption}
\usepackage{subcaption}
\usepackage{hyperref}
\usepackage{placeins}
\usepackage{longtable}
\usepackage{url}
\usepackage{enumerate}
\usepackage[export]{adjustbox}
\usepackage{float}
\usepackage{xcolor}

\usepackage{seqsplit}

\newcommand{\calL}{\mathcal{L}}

\author{Wenhao Gao}
\affiliation[MIT2]{Department of Chemical Engineering, Massachusetts Institute of Technology, Cambridge, MA 02139, United States}

\author{Priyanka Raghavan}
\affiliation[MIT2]{Department of Chemical Engineering, Massachusetts Institute of Technology, Cambridge, MA 02139, United States}

\author{Ron Shprints}
\affiliation[MIT2]{Department of Chemical Engineering, Massachusetts Institute of Technology, Cambridge, MA 02139, United States}

\author{Connor W. Coley}
\email{ccoley@mit.edu}
\affiliation[MIT2]{Department of Chemical Engineering, Massachusetts Institute of Technology, Cambridge, MA 02139, United States}
\alsoaffiliation[MIT2]{Department of Electrical Engineering and Computer Science, Massachusetts Institute of Technology, Cambridge, MA 02139, United States}

\title{Revealing the Relationship Between Publication Bias and Chemical Reactivity with Contrastive Learning}


\begin{document}
\begin{abstract}


A synthetic method's substrate tolerance and generality are often showcased in a ``substrate scope'' table. However, substrate selection exhibits a frequently discussed publication bias: unsuccessful experiments or low-yielding results are rarely reported. 
In this work, we explore more deeply the relationship between such publication bias and chemical reactivity beyond the simple analysis of yield distributions using  
 a novel neural network training strategy,
\emph{substrate scope contrastive learning}. By treating reported substrates as positive samples and non-reported substrates as negative samples, our contrastive learning strategy teaches a model to group molecules within a numerical embedding space, based on historical trends in published substrate scope tables. 
Training on 20,798 aryl halides in the CAS Content Collection\textsuperscript{TM}, spanning thousands of publications from 2010-2015, we demonstrate that the learned embeddings exhibit a correlation with physical organic reactivity descriptors through both intuitive visualizations and quantitative regression analyses. Additionally, these embeddings are applicable to various reaction modeling tasks like yield prediction and regioselectivity prediction, underscoring the potential to use historical reaction data as a pre-training task. 
This work not only presents a chemistry-specific machine learning training strategy to learn from literature data in a new way, but also represents a unique approach to uncover trends in chemical reactivity reflected by trends in substrate selection in publications.

\end{abstract}


\section{Introduction}

\begin{figure}[t!]
    \centering
    \includegraphics[width=0.9\textwidth]{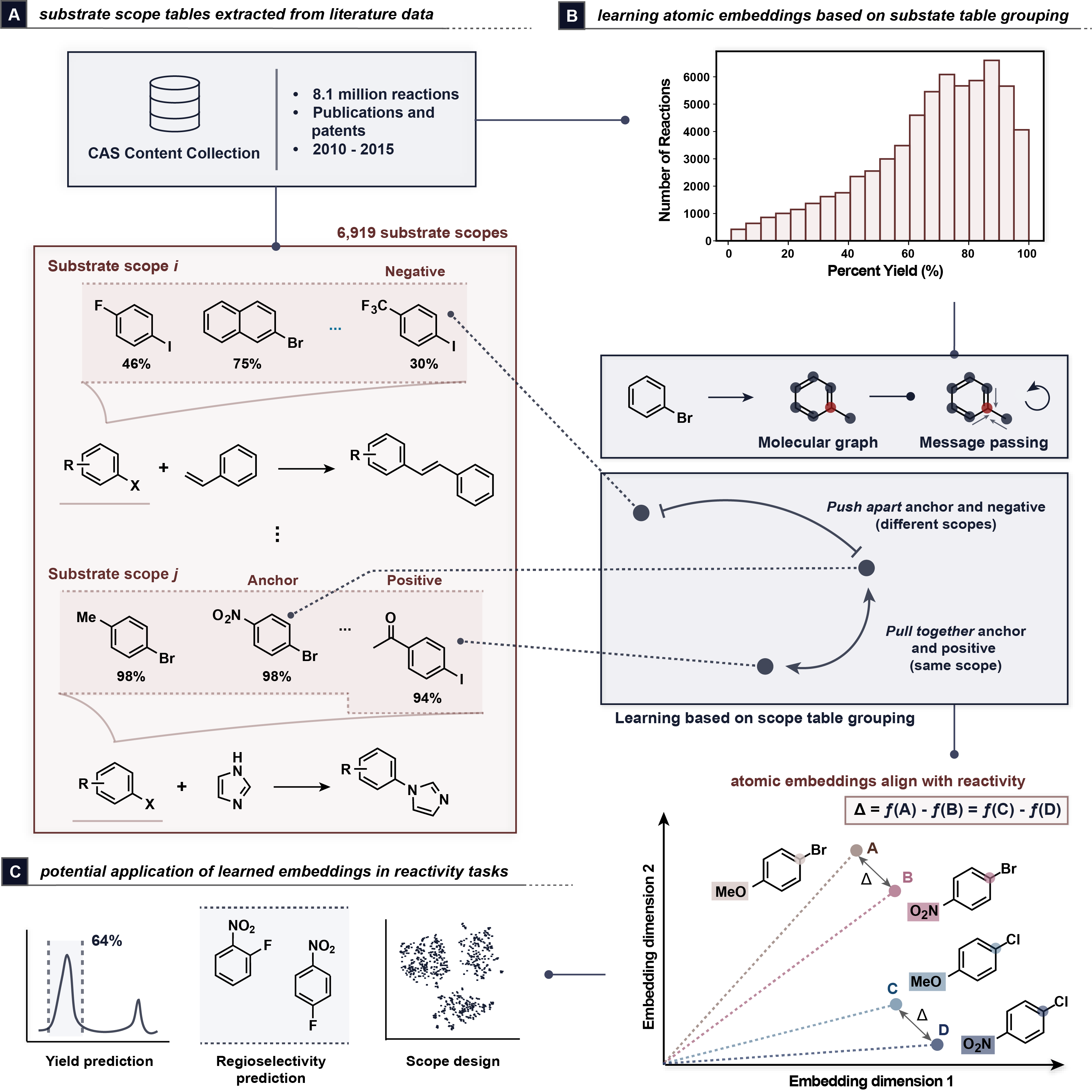}
    \caption{(A) Substrate scope tables for training the network are curated from the CAS Content Collection, focusing on aryl halides with their associated yields. The two substrate scopes $i$ and $j$ shown in the figure are real samples from the database\cite{guastavino2014room,movahed2014one}. (B) Overview of \emph{substrate scope contrastive learning}: The hypothesis behind our training strategy is that publication bias in chemical reactions reveals more subtle reactivity trends than the pronounced inclination towards higher yields as depicted in the histogram. A message-passing neural network operates on molecular graphs to derive atomic embeddings. Embeddings of substrates from the same scope table are pulled together, while embeddings of substrates from distinct scope tables are forced apart. 
    The training aims to provide atomic embeddings that inherently reflect the grouping of reported substrate scopes table, which creates a vector space showcased to be aligned with reactivity trends. (C) The embeddings obtained after contrastive learning can then be analyzed to understand what has been learned and/or used as representations of molecules in various reactivity-related tasks.}
    \label{fig:schematic}
\end{figure}

The report of a new reaction is concurrent with experiments showcasing the scope and limitations of the reaction. These insights are typically derived from evaluating a diverse array of substrates, typically under a constant or similar set of conditions, with results presented in substrate scope tables. 
However, the information presented in literature-reported substrate scope tables exhibits biases (see Figure \ref{fig:schematic}B, with additional statistics in Figure \ref{fig:stat_of_dataset}) and is subject to various sociological factors \cite{kozlowski2022topic,rana2024standardizing,maloney2023negative,williams2023branched}. This bias arises from two primary sources: selection bias, where chemists may choose substrates that are readily available or expected to give favorable results, and reporting bias, where high-yielding ``successful'' outcomes are more likely to be highlighted over low-yielding ``negative'' outcomes. These two factors interact synergistically, collectively termed publication bias\cite{rana2024standardizing}. While publication bias is pervasive across scientific disciplines\cite{fanelli2017meta}, its study within the context of chemistry, especially its relationship with chemical reactivity, remains in its early stages\cite{maloney2023negative}.


Machine learning (ML) models are increasingly employed to model chemical reactivity quantitatively\cite{hickey2015predicting,santiago2018predictive,kariofillis2022using,tang2023interrogating}, opening up the possibility of studying reactivity beyond reported substrates. However, publication bias has led to a disproportionately low amount of negative data, making it difficult for ML models to reliably generalize to unseen compounds\cite{strieth2022machine,maloney2023negative}. 
In this paper, we approach the problem from a different perspective: we design a new training strategy leveraging the publication bias in substrate scope design as a \emph{premise}, train a machine learning model that naturally reflects these historical trends, and examine its correlation with chemical reactivity. Specifically, we introduce \textbf{substrate scope contrastive learning} (ContraScope), a training strategy whose core premise is that substrates reported together in a substrate scope table (with similar yields) are more similar in reactivity than those not reported together on average. Using contrastive learning,\cite{schultz2003learning,hoffer2015deep,khosla2020supervised} we train a graph neural network model to encode reactive atoms into numerical embeddings, ensuring positive data are grouped closely while being distinguished from negative data.
To account for variations within the same substrate scope table, we adjust the embedding distances among positive samples based on their reported yields under consistent conditions. 



We demonstrate our contrastive learning approach using aryl halides, a popular class of substrates employed in a wide range of chemical reactions. Specifically, we curated substrate scope tables from the literature published from 2010 to 2015, as indexed in the CAS Content Collection\textsuperscript{TM}, the largest human-curated collection of scientific data in the world, and demonstrate that the pattern of substrate usage in publications actually captures key aspects of aryl halide reactivity. The learned embeddings show both qualitative and quantitative correlations with traditional reactivity descriptors. Our results show that the publication bias in substrate scope tables, though previously considered detrimental to model training, actually offers valuable insights into reactivity patterns and can serve as a source of information. We hope to encourage further study of the relationship between publication bias and chemical reactivity, as well as the development of additional training strategies that effectively capture chemical reactivity---to further the pursuit of quantitative data-driven modeling in synthetic organic chemistry.

\section{Substrate Scope Contrastive Learning}

The core idea of substrate scope contrastive learning (ContraScope) is to treat substrates reported within the same scope tables as positive samples while randomly sampling molecules that are \emph{not} listed in that published substrate scope table as negative samples. By training the model this way, the resulting embeddings naturally reflect the selective design of substrate scope tables, and we can then investigate them to understand what trends have been learned. We emphasize that this approach does not assume every molecule excluded from a published scope table is incompatible with that reaction; rather, it uses the grouping of literature-reported substrate scope tables as a premise and source of inspiration. If there is some degree of publication bias--i.e., if the excluded substrates are, \emph{from a probabilistic perspective}, less reactive or compatible than those included--then our learned embeddings should capture this underlying information about chemical reactivity.



Specifically, ContraScope trains the model on triplets of molecules: for each molecule as an anchor, we randomly select another molecule from the same substrate scope table to serve as a positive sample, encouraging their embeddings to be closely aligned; concurrently, we sample a molecule that doesn't belong to the reported scope table to act as a negative sample, encouraging their embeddings to be distantly separated. Recognizing that molecules within the same reported substrate scope table can exhibit varying degrees of reactivity, we further encourage the distance between the anchor and positive molecules to be proportional to the difference in their yields. Formally, we train our model by minimizing the following loss function:
\begin{equation}
    \calL (m_a, m_p, m_n) = \underbrace{(d(m_a, m_p) - \gamma |y_a - y_p|)^2}_\text{Distance proportional to yield difference} + \underbrace{\beta \log [1+\exp(M - d(m_a, m_n))]}_\text{Distance at least a margin M away}
\label{eq:loss}
\end{equation}
where $m_a, m_p, m_n$ denote a triplet of molecules, comprising an anchor, a positive, and a negative sample. $d(\cdot,\cdot)$ denotes the embedding distance between two molecules as learned by the model.
$y_a$ and $y_p$ represent the reported yield under identical reaction conditions for the anchor and positive molecules. The margin $M$, alongside the ratios $\beta$ and $\gamma$ are constants determined through hyperparameter tuning (see section \textit{Learning curves and hyper-parameter tuning} in SI for detailed values). These terms ``pull'' substrates from the same substrate scope closer based on how similar their yields are, while they ``push'' away all other aryl halides, on average, not reported in the scope.

Substrate scope tables and associated reaction yields are sourced from a subset of the CAS Content Collection describing publications and patents between 2010 and 2015. We restrict our analysis to aryl halides as they are an important category of molecules widely employed as building blocks in medicinal chemistry \cite{roughley2011medicinal,brown2016analysis}. 
Within each substrate scope,  we verified that all reactions involve transformations at a C--X bond and are performed under identical conditions, including reactants besides the aryl halides in the case of multi-component reactions, according to the database. 
We excluded scopes with fewer than five reactions, resulting in a total of 6,919 scopes with 64,192 reactions. For each training epoch, each aryl halide serves as the anchor in 16 triplets. 
We adopt a graph isomorphism network (GIN)\cite{xu2018powerful}, a widely used graph neural network, as the model architecture. Taking a single substrate at a time as input, the model represents aryl halides as graphs with atom and bond identities as input features\cite{yang2019analyzing}. Details on our training methodology are in the Methods Section.


\section{Results}

\subsection{Visualization and intuitive investigation of the learned aryl halide embeddings} 


\begin{figure}[t!]
    \centering
    \includegraphics[max width=1.00\textwidth]{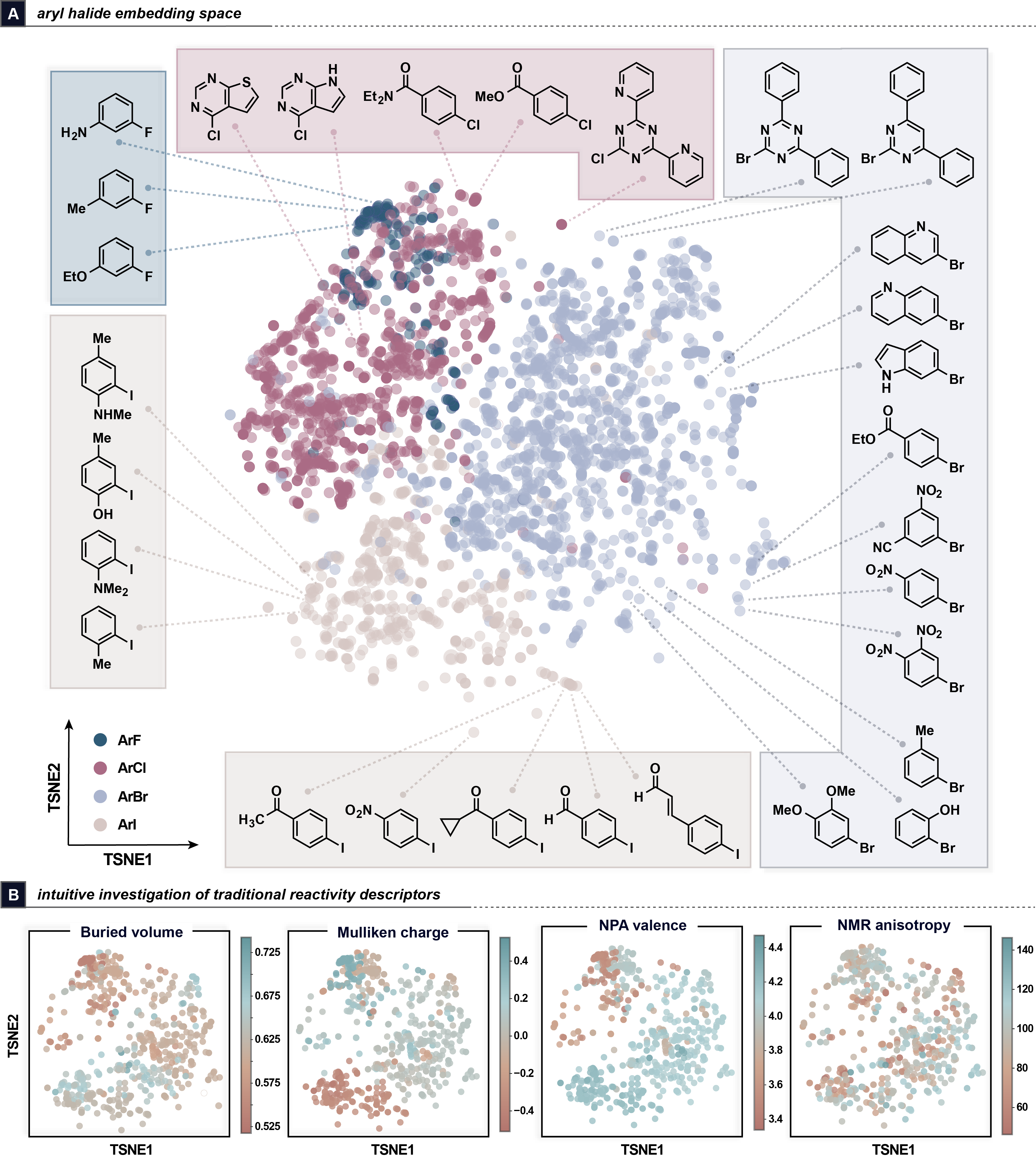}
    \caption{(A) A two-dimensional projection of learned embeddings using t-SNE\cite{hinton2002stochastic} for intuitive investigation of aryl halide embeddings obtained through substrate scope contrastive learning; each point denotes an aryl halide and is colored by halide type. Neighborhoods of similar aryl halides in the embedding space are annotated with structures. (B) t-SNE visualizations of the learned embeddings, colored by traditional reactivity descriptor values.}
    \label{fig:visualization}
\end{figure}

We first visualize the learned embeddings to illustrate how the model has learned to organize the chemical space of aryl halides. To do this, we encode a set of molecules comprising the 500 most frequently used aryl halides, augmented by a random sample of 2922 aryl halides from the substrate scope dataset. Upon encoding, we use t-distributed stochastic neighbor embedding (t-SNE) \cite{hinton2002stochastic} projection to visualize the 64-dimensional embedding learned by the model in Figure \ref{fig:visualization}A (see Figure \ref{fig:bef_aft} for results of other projection methods and a comparison with embeddings from an untrained network). 

A closer look at the position of specific structures in the embedding space (referred to as call-outs in Figure \ref{fig:visualization}A) reveals that molecules with qualitatively similar reactivities---characterized by either electron-withdrawing groups (e.g., nitro, aldehyde, carbonyl) or electron-donating groups (e.g., hydroxyl, ether, amine, alkyl)---tend to cluster together. This pattern is consistent across various halide classes, underscoring the broad applicability of the embeddings we have developed. The pairwise distances of embeddings of 12 para-substituted bromobenzenes are visualized in Figure \ref{fig:fig_pair_wise_dist}, revealing closer clustering for electron-withdrawing groups (e.g., nitrile, ester) compared to electron-donating groups (e.g., methoxy, amine). This trend, absent in Tanimoto distances from Morgan fingerprints (Figure \ref{fig:fig_pair_wise_dist}), underscores the embedding space's superior ability to capture functional reactivity similarities.

To further investigate the connection between our data-driven, substrate scope-informed embeddings and chemical reactivity, we calculated and visualized established reactivity descriptors\cite{santiago2018predictive} calculated with first-principle quantum calculations\cite{kohn1996density} in the embedding space. We encode a set of 762 aryl halides comprising of 500 of the most frequently appearing aryl halides from our dataset and 262 monosubstituted aryl halides with functional groups sourced from a Hammett constant table\cite{hansch1991survey}. We calculated reactivity descriptors using Gaussian 16\cite{g16} and auto-qchem\cite{zuranski2022auto}, with a particular focus on local atomic properties at the aromatic carbon bonded to the halogen, as our contrastive learning approach is designed to learn reactivity localized to the C--X motif, rather than molecule-level properties. Subsequent t-SNE projection mapping of this aryl halide set, as illustrated in Figure \ref{fig:visualization}B, revealed that molecules positioned closely in the embedding space exhibited similar values of these atom-level reactivity descriptors. While the strongest differentiator is the identity of the halogen, more subtle patterns within each halide class can be seen.


\subsection{Regression analysis reveals the relationship between the learned embeddings and reactivity descriptors} 

\begin{figure}[t!]
    \centering
    \includegraphics[max width=0.95\textwidth]{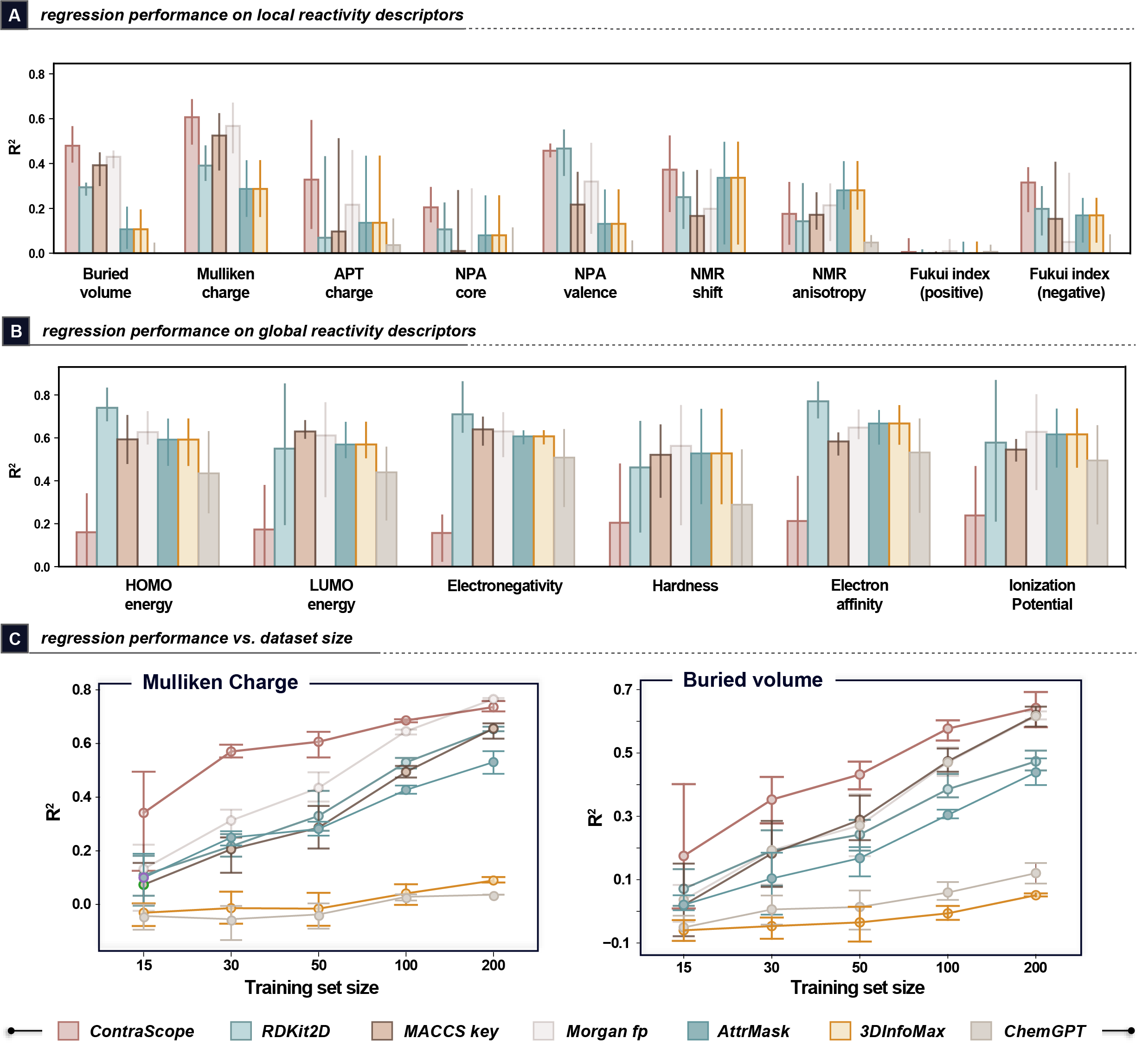}
    \caption{Relationship between the learned embeddings and conventional reactivity descriptors. 
    (A and B) Regression performance ($r^2$) when learning to predict local (B) and global (C) physical organic chemistry descriptors from different input representations using linear models; negative values are excluded from the plot. 
    (C) Analysis of support vector machine (SVM)\cite{hearst1998support} regression performance as a function of dataset size, highlighting the efficacy of our embeddings in scenarios with limited data. }
    \label{fig:correlation}
\end{figure}

To quantitatively assess how our learned embeddings capture chemical reactivity, we performed a regression analysis using simple (linear) machine learning models to predict various computed reactivity descriptors from the ContraScope embeddings. This approach allows us to evaluate the correlation between the learned pattern in our multidimensional embedding and descriptors we know to be relevant to chemical reactivity. By restricting our approach to simple linear models, we limit the overall modeling capacity; thus, strong regression performance would be driven primarily by the intrinsic relationship between the embedding features and the target descriptors. This would therefore gauge the extent to which publication trends alone reveal trends in chemical reactivity. We compare model performance using our embeddings, widely-used molecular featurizations\cite{ashton2002identification} or fingerprints\cite{rogers2010extended}, as well as embeddings from three representative pre-trained deep learning models: (1) the same GIN architecture pre-trained using attribute masking (AttrMask)\cite{hu2020strategies}, (2) a model pre-trained to capture three-dimensional molecular conformations (3DInfoMax)\cite{stark20223d}, and (3) a state-of-the-art chemical language model (ChemGPT)\cite{frey2023neural}.

We calculated commonly used reactivity descriptors for the 762 aryl halides discussed in the previous section and performed regression analysis on them. Our learned embeddings outperformed or at least be comparable to other featurization and pre-trained embeddings in predicting most local reactivity descriptors related to the C–X motif at the carbon atom (Figure~\ref{fig:correlation}A), with the clearest advantages observed for partial charge-related descriptors and buried volume. Notably, as shown in Figure \ref{fig:correlation}C, this performance gap becomes more significant when fewer data points are available, demonstrating that our pre-training procedure provides informed embeddings as a robust initialization that markedly improves downstream modeling performance, indicating the close relationship between learned representations and both Mulliken charges and buried volumes. As the training set size increases, we observe the apparent advantage in representation to deteriorate and expect the advantage to disappear entirely at even larger training set sizes.
In contrast, correlations were weaker for global (molecule-level) reactivity descriptors (Figure~\ref{fig:correlation}B), which is expected due to the design of our pre-training strategy focusing on reactivity localized to the C–X motif. Overall, these findings confirm that the multi-dimensional patterns learned from the historical trend in substrate scope grouping strongly correlate with local reactivity characteristics while exhibiting a lower correlation with global reactivity descriptors.





\subsection{Potential downstream application of the learned embeddings}

\begin{figure}[t!]
    \centering
    \includegraphics[max width=0.95\textwidth]{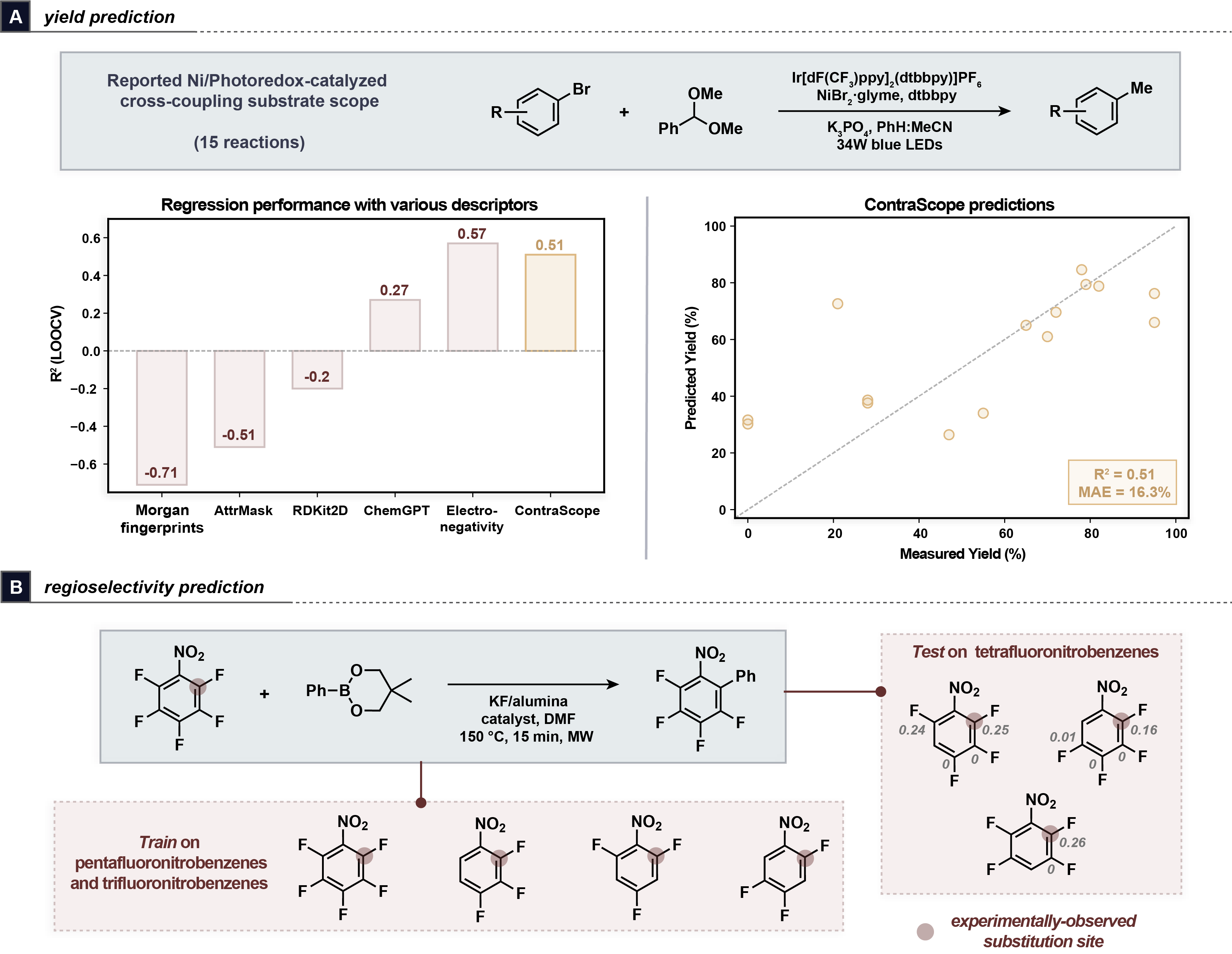}
    \caption{Validation and application of learned embeddings in various downstream tasks. (A) Application of ContraScope embeddings to predict the yield of aryl bromides in cross-coupling reactions\cite{kariofillis2022using}. We show its predictive performance via leave-one-out validation compared with other common featurizations and a scatter plot mapping our predictions against experimental yields.  (B) Application of ContraScope embeddings to predict the regioselectivity in arylation reactions of fluorobenzenes, with the model trained on penta- and trifluoronitrobenzenes to anticipate reactivity in tetrafluoronitrobenzenes, highlighting the reaction centers confirmed by experiments and annotated with prediction outcomes\cite{cargill2010palladium}. 
    }
    \label{fig:application}
\end{figure}

In this section, we showcase potential downstream applications of our learned embeddings for reaction performance prediction tasks. These evaluations serve to demonstrate how information relevant to the downstream tasks is implicitly revealed by publication bias. If it were not, we would expect ContraScope embeddings to be unable to perform these tasks. We do not aim to or claim to achieve state-of-the-art performance on supervised learning tasks of reaction performance.


\textbf{Yield prediction.} As a first case study, we examine the prediction of reaction yields of various aryl bromides under identical reaction conditions. We select a Ni/photoredox catalyzed cross-coupling reaction, with reported yields for a diverse set of substrates published after 2015\cite{kariofillis2022using}, and therefore not contained in our pre-training set. \citeauthor{kariofillis2022using} report achieving a validation $r^2$ of $0.57$ with a univariate model based on a computed DFT descriptor, specifically electronegativity. Without being trained on any physical organic chemistry concepts or features (like electronegativity) and without requiring the cost of electronic structure calculations, our learned ContraScope embedding is able to achieve a comparable $r^2$ of 0.51 (Figure \ref{fig:application}A). Other common fixed or pre-trained representations based on structure are far less successful, further confirming that trends in substrate scope publication contain information relevant to reactivity.


\textbf{Regioselectivity prediction.} As a second case study, we applied our embeddings to compare the likelihood of multiple reactive sites within a molecule. We focused on a palladium-catalyzed Suzuki-Miyaura coupling reaction using polyfluoronitrobenzenes \cite{cargill2010palladium}. The model was trained with yield data from four penta- and trifluoronitrobenzenes, and used to predict the reactivity of different C-F sites in three tetrafluoronitrobenzenes. The results, quantitatively confirmed by experimental data, show that a support vector machine (SVM) model using our aryl halide representations correctly learns to predict the expected reactivity at the C--F site \emph{ortho} to the nitro group, but also distinguishes subtle differences in electronic effects, particularly where two adjacent C--F sites are nonequivalent (Figure \ref{fig:application}B). This result highlights that ContraScope's atomic-level representations can distinguish similar reactive sites and underscores the importance of learning atomic embeddings rather than molecular embeddings for reactivity prediction tasks. 


\section{Discussion}


Substrate scope tables from journal articles, commonly remarked upon as being being biased towards high-yielding examples, are, in fact, biased in subtler ways that correlate with chemical reactivity. 
By learning the patterns of what substrates were included or excluded in a publication, ContraScope extracts the underlying expert knowledge and reactivity trends used (implicitly or explicitly) to make that decision. 
We illustrate this both qualitatively and quantitatively, showing how our embeddings--trained only on substrate groupings in publications--align with trends in reactivity according to computed descriptors and expert intuition. 
To our knowledge, this is the first approach that studies publication bias as a source of information, rather than as a hindrance to modeling.



We would like to emphasize that publication bias serves  as the premise and inspiration for designing the loss function without claiming or requiring that all molecules outside the reported scope tables are necessarily unreactive. Our premise that all molecules not reported in a substrate scope table are less similar to those that are reported is a simplification that holds true only on average. We acknowledge that this approach inherently introduces some false negatives during training--molecules that could undergo the reaction but are treated as negative samples. However, it is well established that machine learning models can still extract meaningful insights and achieve reasonable performance despite the presence of noisy or mislabeled data points. In our case, the model still finds signal in the patterns of substrates included in the reported scope tables that uncovers implicit reactivity trends. While excluding false negative samples through expert curation may enhance the model’s quantitative performance, addressing this challenge without conflating trends from publication bias with trends from such curation bias is complex and falls beyond the scope of this work, though some exploratory trials are included in the Supporting Information.

The downstream applications showcased in this paper---such as the supervised learning of reaction yields---show the versatility and potential value of the learned ContraScope embeddings only as proofs-of-concept; the learned embedding approach, like any other, is still not a universal solution to molecular representation learning.
For instance, a nearest neighbor regression analysis of yield across all training substrate scopes indicates that two molecules with similar embeddings do not necessarily exhibit similar yields, which is similar to other common features (Figure \ref{fig:cas_r2_dist}). Moreover, substrates may achieve dissimilar reaction yields as a result of side reactions rather than inherent dissimilarity in reactivity at the C--X motif, which would not be captured in our approach; this may partially explain the observed lack of correlation for structurally complex aryl halides in the chemistry informer library \cite{kutchukian2016chemistry} (Figure \ref{fig:Fig_informer}). 
From a practical standpoint, our loss function combines a term for anchor-positive loss with one for anchor-negative loss in a linear manner, which 
can, at times, lead to instability in the training process.
All of these challenges underscore that understanding and predicting chemical reactivity remains an open question, highlighting the need for further refinement of training strategies and loss functions specially designed for chemical data.

Overall, our work presents a conceptual framework for studying publication bias in a unique way through the lens of deep representation learning. 
We show that the resulting learned embeddings correlate with physical organic reactivity descriptors and experimental reaction performance as evidence that this information is implicit in patterns of substrate usage. This methodology can be straightforwardly extended to study trends in additional substrate classes beyond aryl halides by changing the training dataset. We are optimistic that our framework and open-source code will serve as a catalyst for further advancements in the field, fostering more functionality-aligned training strategies in molecular representation learning and the quantitative study of publication bias.


\section{Methods} \label{sec:methods}

\noindent \textbf{Substrate data for training} In this study, we focus on aryl halides. We partitioned reactions recorded in the CAS Content Collection between 2010 and 2015 into substrate scope tables comprised of reactions wherein the only variable is a single aryl halide substrate. All other substrates and recorded conditions remained constant within each scope. Importantly, all reactions take place at an aryl C--X bond. Scopes with fewer than five reactions and reactions without recorded yields were excluded. This led to our final training dataset of 20,798 distinct aryl halides, covering 64,192 reactions and 6,919 substrate scopes. 

\noindent \textbf{Calculation of reactivity descriptors} All reactivity descriptors are calculated by density-functional theory (DFT) with Gaussian 16\cite{g16} via an automated descriptor generation pipeline built on top of AutoQChem\cite{zuranski2022auto}. For each unique aryl halide, up to 20 conformers were generated using RDKit's ETKDG algorithm\cite{riniker2015better} and optimized using the MMFF94 force field\cite{halgren1996merck}. To reduce computational overhead when using DFT, the lowest-energy conformer for each molecule was then selected via GFN2-xTB\cite{bannwarth2019gfn2}. Any conformer for which any energy calculation did not converge was discarded. The lowest-energy conformer for each aryl halide then underwent geometry optimization and frequency calculations in Gaussian 16 using the B3LYP functional\cite{becke1992density,lee1988development} with the 6-31G*\cite{petersson1988complete} basis set. Atoms  with atomic number $>35$ use the LANL2DZ\cite{hay1985ab} basis set instead. This led to the generation of 25 molecule-level descriptors per molecule (energies, energy corrections, dipole moment, HOMO/LUMO energies, electronegativity, etc.) and 19 atom-level descriptors per atom per molecule (buried volume, partial charges, NMR shielding constants, etc.), and the relevant atom-level descriptors for the reactive C--X bond were extracted. The full set of descriptors is listed in Figures~\ref{fig:global_desc} and \ref{fig:local_desc}.

\noindent \textbf{Featurization and network architecture} We adopt a graph representation of molecules and the graph isomorphism network (GIN)\cite{xu2018powerful} for modeling them. GIN is one kind of graph neural network that updates the representation of each atom over multiple iterations as follows:
\begin{equation}
    h_v^{(k)} = \text{MLP}^{(k)}\left((1+\epsilon^{(k)})h_v^{(k-1)}+\sum_{u \in \mathcal{N}(v)}h_u^{(k-1)} \right)
\end{equation}
where $h_v^{(k)}$ is the atom representation of atom $v$ at $k$-th layer, $\mathcal{N}(v)$ is a set of atoms that are covalently bonded to atom $v$. $\epsilon$ is a learnable scalar and $\text{MLP}$ represents a multi-layer preceptrons model. Utilizing this atom-wise message-passing schema, GIN is able to propagate and aggregate information through network layers to embed atoms in a manner that respects graph isomorphism. After hyper-parameter tuning, the network has a hidden dimension of 64, and message passing layer of 5. For the initial representation of molecular graphs, we use minimal featurizations\cite{yang2019analyzing} that include atomic numbers, total degrees, formal charges, chiral tags, number of hydrogens, hybridization types, aromaticity, mass, and bond types.

\noindent \textbf{Substrate scope contrastive learning} The training of substrate scope contrastive learning involves triplet sampling, computing loss function depicted in Eq. \ref{eq:loss}, and back-propagation. Within an epoch, each aryl halide serves as the anchor molecule once and 16 triplets are sampled randomly for each anchor. We exclude all cases with identical molecules sampled in one triplet. 
Based on empirical performance, we adopt a distance metric called signal-noise ratio distance (SNR) \cite{yuan2019signal}, defined as follows:
\begin{equation}
    d(m_i, m_j) = d_{\text{SNR}}({f}_i, {f}_j) = \frac{var({f}_i)}{var({f}_j - {f}_i)}
\end{equation}
where ${f}_i, {f}_j$ is the embedding of aryl halide $m_i, m_j$. We take the representation of an aryl halide's reactivity, $f_i$, to be the atom-level feature of the carbon atom in the reactive C--X bond from the GIN. The Adam\cite{kingma2014adam} optimization algorithm is used for stochastic gradient descent. 

To monitor the model performance during the training, we collected four datasets evaluating the ability of the learned embeddings to adapt to downstream tasks: 15 aryl bromides with associated reaction yields for a cross-coupling reaction\cite{kariofillis2022using}, 61 mono-substituted aryl halides with Hammett constants\cite{hammett1937effect}, and calculated Muliken charges and NMR shielding constants. At the end of each epoch, we used the learned embeddings as features, selected the most informative features indicated by mutual information\cite{shannon1948mathematical} and the most correlated features indicated by the Pearson correlation coefficient, and trained a k-Nearest Neighbor (k-NN) to predict the reaction yields, Hammett constants, Muliken charges and NMR shielding constants. We evaluated the $r^2$ \cite{wright1921correlation} of prediction with leave-one-out cross-validation on the four validation datasets and summed them up as a single scalar value for monitoring the training process and determining the time of early termination of the training. Training curves are shown in Figure \ref{fig:curves}.


\noindent \textbf{Regression analysis with traditional reactivity descriptors} 
To validate if the embedding learns anything about reactivity, we selected the most commonly used 500 aryl halides from the training set and supplemented them with p-, m-, and o-substituted aryl halides with common functional groups selected from a study of Hammett constants\cite{hansch1991survey}, providing a dataset with 762 aryl halides in total. Reactivity descriptors were calculated for this set of 762 molecules. For each molecule, we computed RDKit2D descriptors\cite{landrum2013rdkit} and Morgan fingerprints\cite{rogers2010extended} (1024 bits, radius 2) using Therapeutic Data Commons (TDC)\cite{huang2021therapeutics}. As for pre-trained representation baselines, we adopted AttrMask, 3DInfoMax, and ChemGPT. Among them, AttrMaskis a pre-trained GIN with node-masking\cite{hu2020strategies} and ChemGPT\cite{frey2023neural} is a chemical language model with 4.7 M parameters, both available from Deep Graph Library (DGL)\cite{wang2019deep} and HuggingFace\cite{wolf2019huggingface}, interfaced via molfeat\cite{molfeat}. 3DInfoMax\cite{stark20223d} is a graph neural network model pre-trained to incorporate information from three-dimensional molecular conformations. In our work, we adopt the pre-trained parameters provided by the original repository to ensure consistency. For Figure \ref{fig:correlation}B, we assessed k-Nearest Neighbor (k-NN), linear regression, L1 and L2 norm, and support vector machine (SVM)\cite{hearst1998support} on each embedding, reporting the best $r^2$ performance. Performance validation used a 3-fold cross-validation, with mean and range of $r^2$ for the top model is shown in the bar plot. In Figure \ref{fig:correlation}C, we trained an SVM model\cite{hearst1998support} on a specific subset of the training data, using the remaining data for performance evaluation. This was repeated thrice with independent seeds, reporting the mean and range. For all machine learning models, we adopt the implementation from scikit-learn\cite{scikitlearn}.

\noindent \textbf{Downstream applications} For yield prediction, we used 15 aryl bromides with experimental yields\cite{kariofillis2022using}, employing leave-one-out validation. Molecular embeddings were calculated by the ContraScope-trained GIN as previously described, and other common featurization was calculated using the same pipeline as above. For all embeddings, we selected features that are the most informative or most correlated with the prediction targets indicated by mutual information\cite{shannon1948mathematical} and Pearson correlation coefficient, and tested models including k-NN, linear regression, L1 and L2 norms, SVM, random forest, and MLP, reported the highest validation $r^2$. The $r^2$ value of the univariate model using electronegativity is from the original study\cite{kariofillis2022using} for comparison. For regioselectivity prediction, training data comprised penta- and trifluoronitrobenzenes to predict reactivity at C-F sites in tetrafluoronitrobenzenes\cite{cargill2010palladium}. Each C--F site was encoded and labeled with its yield if reactive, or zero otherwise. During inference, all C--F sites were encoded, and the model predicted each site's reactivity as a binary classification task. For substrate scope design, we used the same purchasable aryl bromides as \citeauthor{kariofillis2022using} and encoded them for K-Means clustering into 15 groups, selecting a representative from each. All machine learning models were implemented using scikit-learn\cite{scikitlearn}.

\section{Supporting Information} \label{sec:SI}

The Supporting Information includes links to the code and validation data, a discussion on the differences between the introduced method and existing pre-training approaches, additional statistics on the training dataset, and further experimental and computational results.

\begin{acknowledgement}

This research was supported by the National Science Foundation under Grant No. CHE-2144153. W.G. is supported by the Google PhD Fellowship. W.G. and P.R. received additional support from the MIT-Takeda Fellowship program. R.S. received additional support from the MIT UROP Office. We thank CAS for providing data from the CAS Content Collection needed to enable this study and Yitong Tseo for preliminary work exploring the dataset.

\end{acknowledgement}





\bibliography{paper}





\newpage
\pagebreak

\renewcommand{\thefigure}{S\arabic{figure}}
\renewcommand{\thetable}{S\arabic{table}}
\setcounter{figure}{0} 
\setcounter{table}{0} 
\setcounter{page}{1}

\begin{centering}

\textsf{\Large{\textbf{Supporting Information}}}

\vspace{0.5cm} 
\textsf{\Large{\textbf{Revealing the Relationship Between Publication Bias and Chemical Reactivity with Contrastive Learning}}}

\vspace{0.3cm}

\textsf{\large{Wenhao Gao,$^\dagger$ Priyanka Raghavan,$^\dagger$ Ron Shprints,$^\dagger$ and Connor W. Coley,$^{\ast,\dagger,\ddagger} $}}
\begin{center}
{\it \small $\dagger$ Department of Chemical Engineering, Massachusetts Institute of Technology, Cambridge, MA 02139, United States}  \\
{\it \small $\ddagger$ Department of Electrical Engineering and Computer Science, Massachusetts Institute of Technology, Cambridge, MA 02139, United States}  \\

\textsf{\small E-mail: ccoley@mit.edu}
\end{center}

\end{centering}

\vspace{0.5cm}

\normalsize
\section{Code and Data Availability}

All code, pretrained model weights, and validation data necessary to reproduce the major results of this work can be found at  \url{https://github.com/wenhao-gao/substrate_scope_contrastive_learning/tree/main}. We are unable to share the full dataset from the CAS Content Collection that contains the full substrate scope information used for contrastive learning.

\section{The misalignment of current pre-training methods and functionalities}

Recent years have witnessed significant advancements in representation learning for molecules, accompanied by a diverse array of pre-training strategies. These strategies have emphasized various aspects, including chemical valence\cite{hu2020strategies}, structural similarity\cite{wang2022molecular}, and conformational information\cite{liu2021pre,stark20223d}. Despite these developments, a consistent or marked enhancement in performance for downstream tasks remains elusive, as noted in Sun et al. (2022)\cite{sun2022does}. A contributing factor to this challenge may be the misalignment between the objectives of these learning models and the specific requirements of their target applications, particularly in modeling molecular functionality.

\begin{figure}[H]
    \centering
    \includegraphics[width=\textwidth]{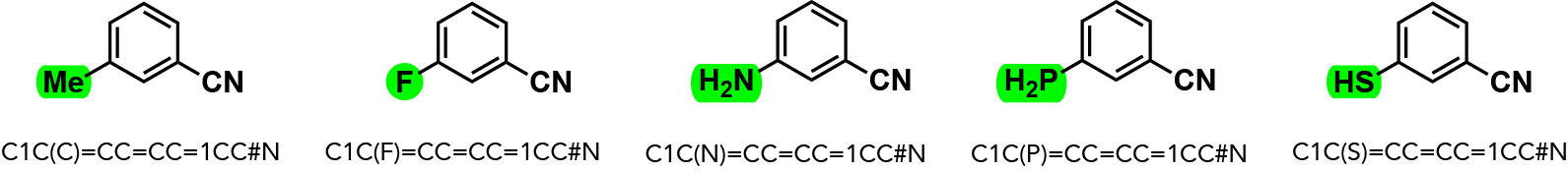}
    \caption{Illustration of various substituted benzonitriles. Despite the structural similarity of the benzonitrile backbone across all molecules, the diverse electronic properties of the substituents significantly influence the molecular functionality. This exemplifies the limitation of some pre-training methods in graph neural networks that may not distinguish between these functional nuances. The corresponding SMILES notations are provided below each molecule, indicating that the highlighted substituents are also interchangeable in string representation, which also brings into question the reasonableness of string-based pre-training methodologies.}
    \label{fig:examples}
\end{figure}

Focusing on graph neural networks, early methodologies primarily targeted node or contextual prediction, as detailed in Hu et al. (2020)\cite{hu2020strategies}. These approaches often classified atoms or groups with identical chemical valences as analogous. This technique, however, shows limitations, as evidenced in the molecules depicted in Figure \ref{fig:examples}. Here, the pre-training method erroneously identifies distinct functional groups as identical, overlooking substantial differences in their chemical functionalities. Similarly, traditional contrastive learning, which views structurally similar molecules as comparable in the embedding space\cite{wang2022molecular}, falls short. The assumption that consistent valence bonds or structural resemblance equates to analogous molecular properties is often flawed, as significant variances may arise.
While the integration of 3D conformational data offers benefits for properties reliant on specific conformations, such as those derived from quantum chemical computations\cite{liu2021pre,stark20223d}, its utility diminishes in broader biological or reaction contexts.
Additionally, language model pre-training that focuses on string representations like SMILES\cite{weininger1988smiles} tends to prioritize syntax comprehension over understanding the intrinsic molecular significance.

This view of the landscape of pre-training approaches for small molecules, intended to be used for the prediction of chemical reactivity but only using information about structure and conformation, served as inspiration for this study. ContraScope is a pre-training approach that is fundamentally based on chemical reactivity.

\section{Statistics of the dataset}

\begin{figure}[H]
     \centering
    \begin{subfigure}[t]{0.32\textwidth}
        \includegraphics[width=\textwidth]
        {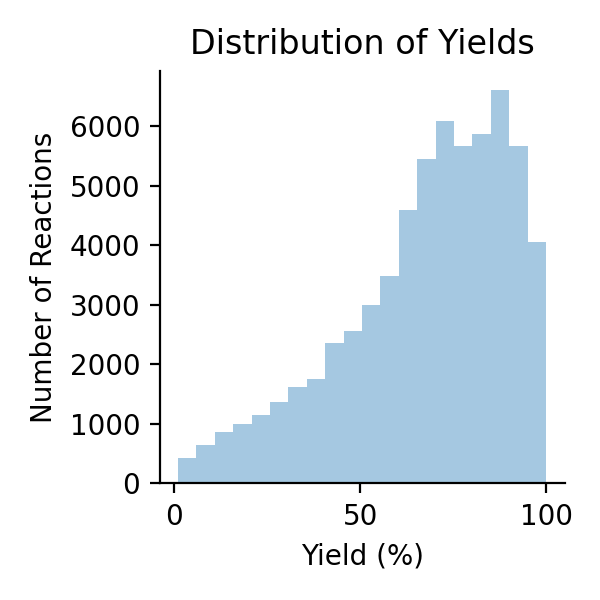}
        \caption{The distribution of the yields of reactions used in training.}
    \end{subfigure}
    \hfill
    \begin{subfigure}[t]{0.32\textwidth}
        \includegraphics[width=\textwidth]        {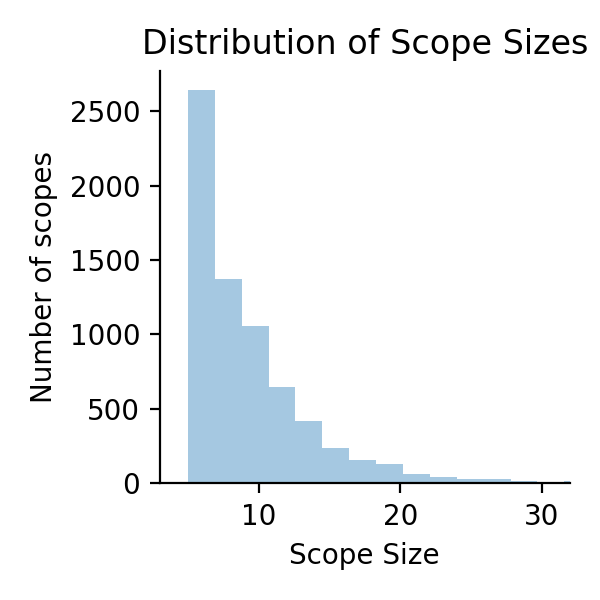}
        \caption{The distribution of the sizes of substrate scopes.}
    \end{subfigure}
    \hfill
    \begin{subfigure}[t]{0.32\textwidth}
        \includegraphics[width=\textwidth]{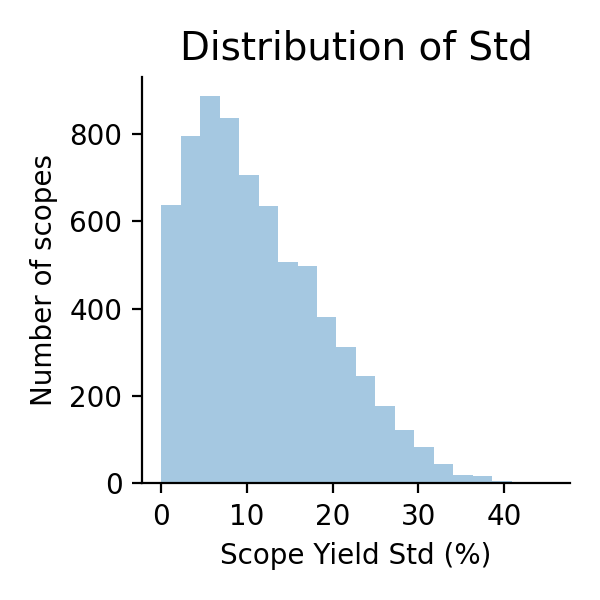}
        \caption{The distribution of the yields' standard deviation within substrate scopes.}
    \end{subfigure}
    \caption{The statistics of the substrate scope dataset used for training.}
    \label{fig:stat_of_dataset}
\end{figure}

In Figure \ref{fig:stat_of_dataset}, we present statistics of the substrate scope dataset we derive from the CAS Content Collection\textsuperscript{TM}, which includes the size distribution of the substrate scopes, the yield values' distribution for all reactions utilized in the training set, and the distribution of the yield standard deviation within these scopes. The data reveal a predominant trend of small substrate scopes, with the majority comprising fewer than 20 substrates. Additionally, there is a discernible skew towards higher yields within the dataset indicating the artificial selective bias that hindered typical supervised machine learning.

\begin{figure}[H]
    \centering
    \includegraphics[width=0.65\textwidth]{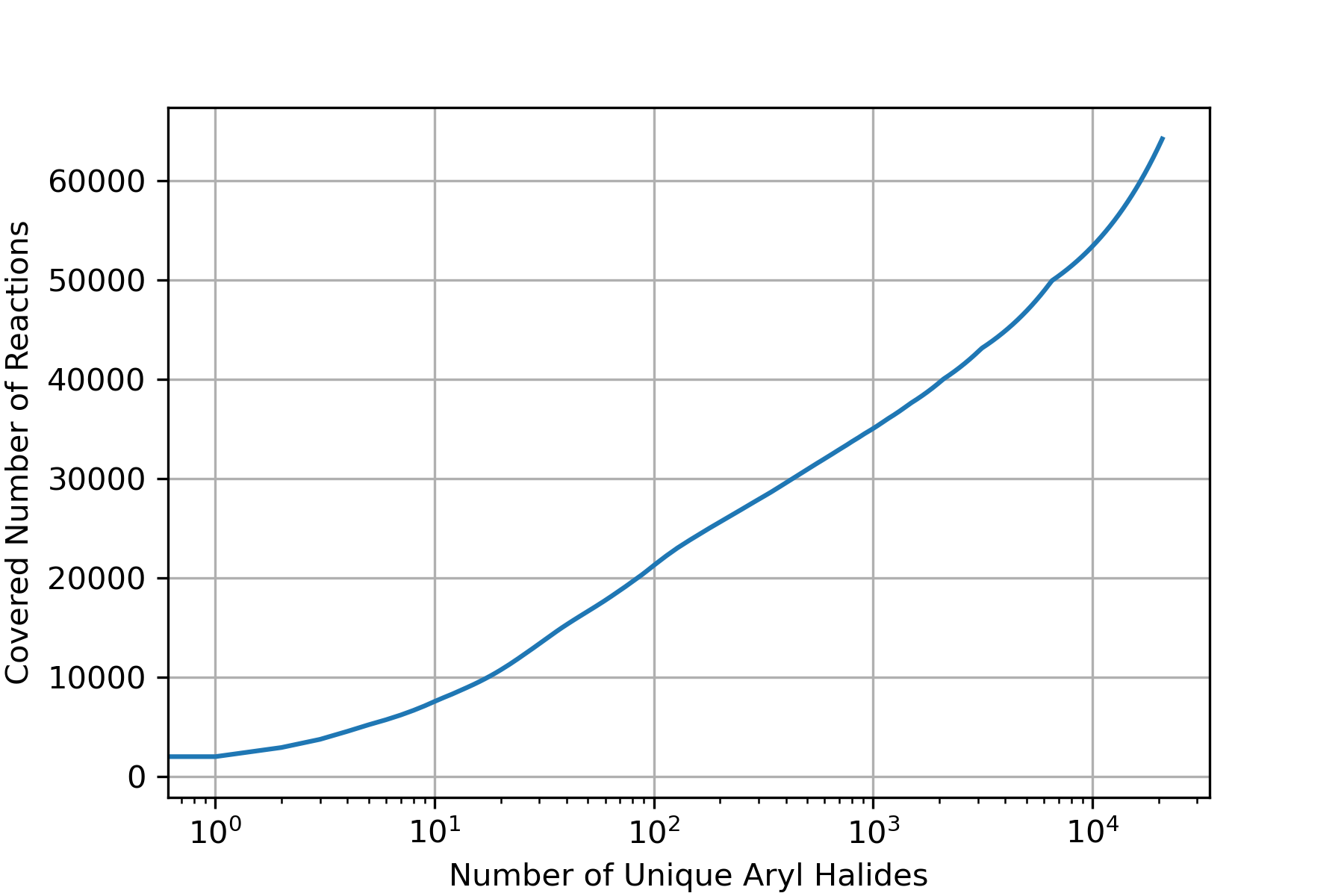}
    \caption{Illustration of the coverage of reactions by the top-k most frequently occurring aryl halides. The x-axis, which is log-scaled for better visual comprehension, represents the k aryl halides ranked by frequency of appearance. It is observed that the 500 most prevalent aryl halides account for over half of the reactions in the dataset.}
    \label{fig:unique_arylhalide_covered}
\end{figure}

In Figure \ref{fig:unique_arylhalide_covered}, we show the cumulative coverage of reactions by the top-k most frequently occurring aryl halides. Notably, the data indicate that the 500 aryl halides with the highest frequency of occurrence account for more than half of the total reactions in the training set. This observation underscores the existence of a frequently utilized subset of aryl halides and substantiates our methodological choice to focus on a select group of these compounds for in-depth analysis.

\begin{figure}[H]
    \centering
    \includegraphics[width=0.65\textwidth]{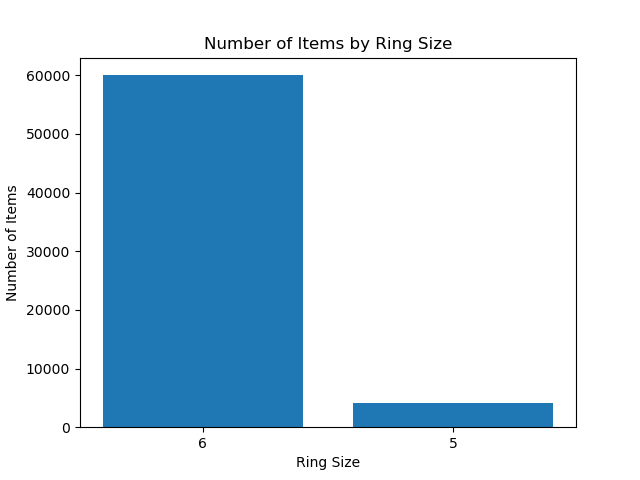}
    \caption{Bar plot showing the distribution of aromatic ring sizes in the training dataset.}
    \label{fig:ring_size}
\end{figure}

\begin{figure}[H]
    \centering
    \includegraphics[width=0.65\textwidth]{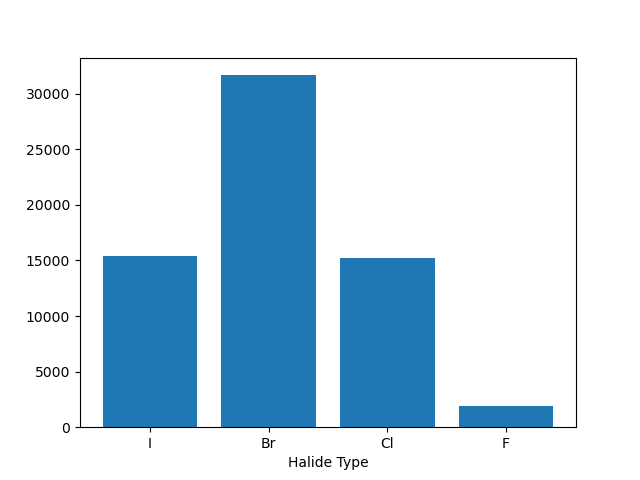}
    \caption{Bar plot illustrating the distribution of different halide types in the training dataset.}
    \label{fig:halide_type}
\end{figure}

\begin{figure}[H]
    \centering
    \includegraphics[width=\textwidth]{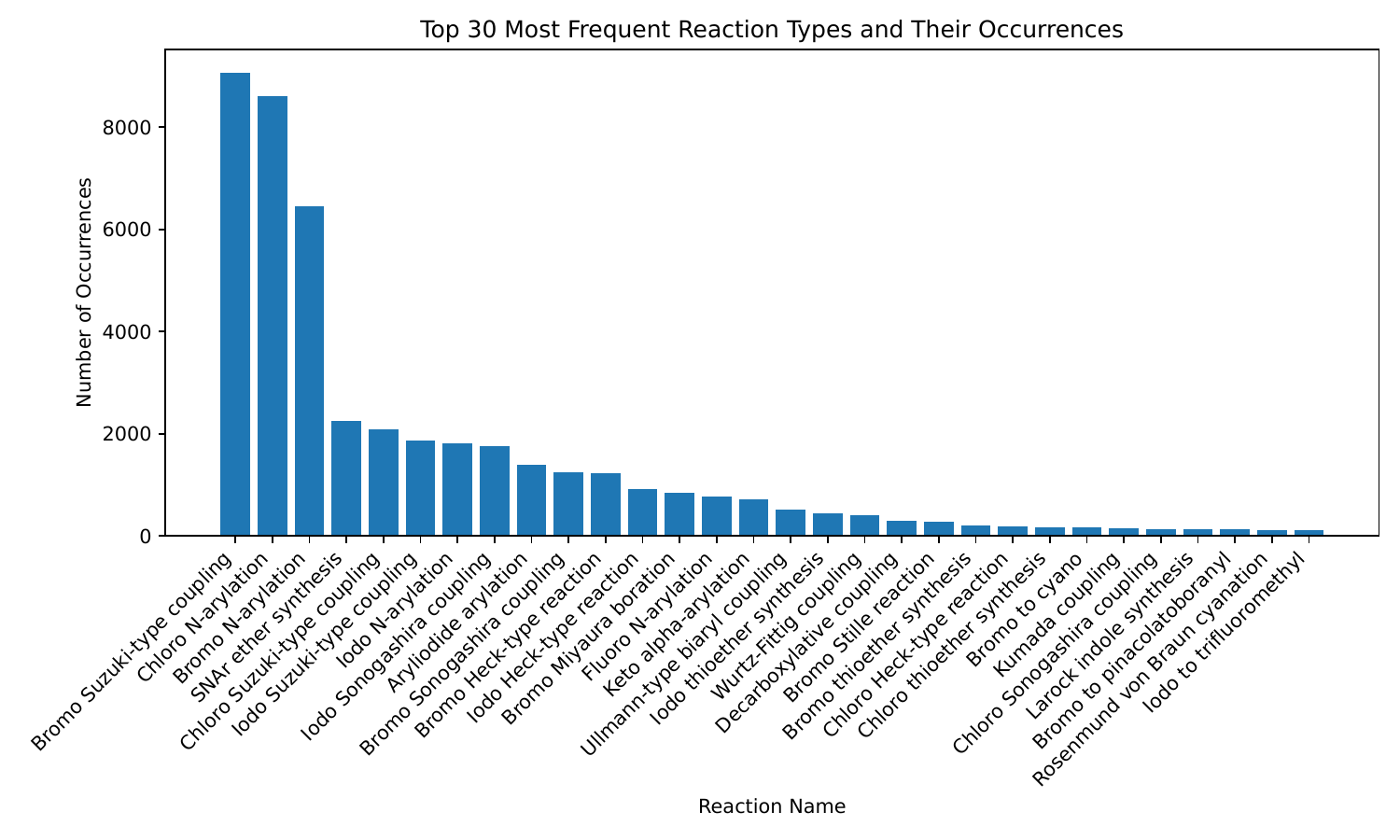}
    \caption{The bar plot displays the 30 reaction types with the number of occurrences in our training data, illustrating their relative prevalence. Each reaction type is labeled on the x-axis, with the number of occurrences represented on the y-axis.}
    \label{fig:top_30_reaction}
\end{figure}

\begin{figure}[H]
    \centering
    \includegraphics[width=0.95\textwidth]{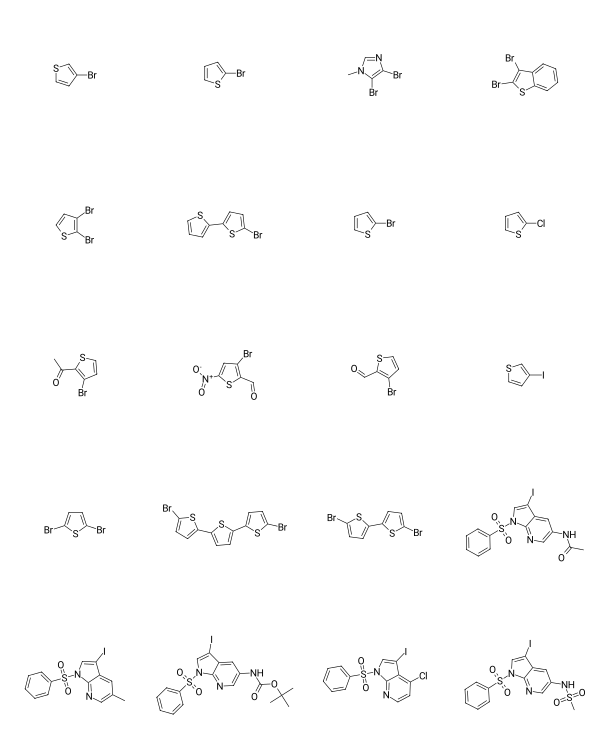}
    \caption{Examples of five-membered aromatic rings present in our training dataset.}
    \label{fig:five_membered_mols}
\end{figure}

\section{Learning curves and hyper-parameter tuning}
\label{sec:curve_tuning}

\begin{figure}[H]
     \centering
    \begin{subfigure}[t]{0.49\textwidth}
        \includegraphics[width=\textwidth]{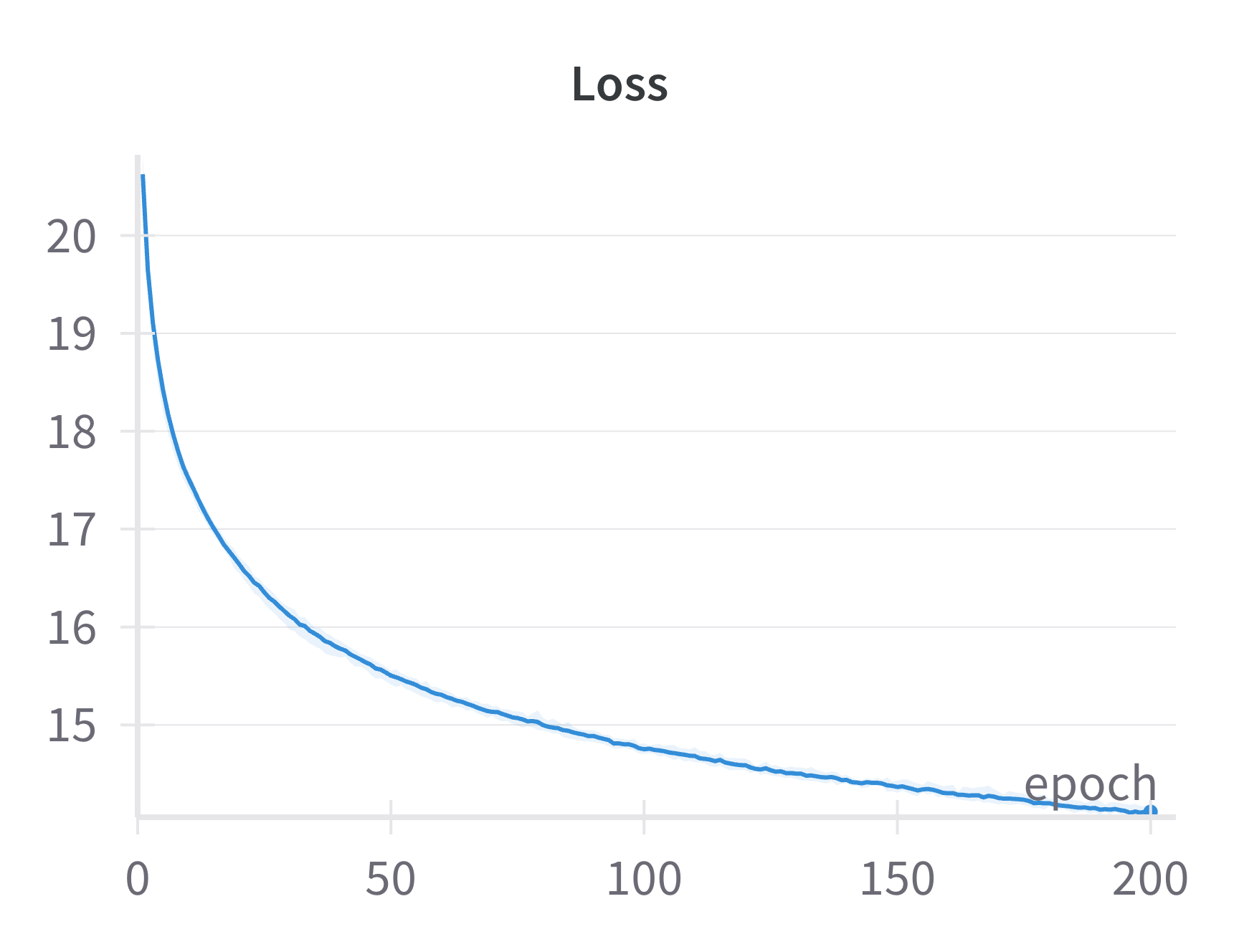}
        \caption{Total substrate scope contrastive loss.}
    \end{subfigure}
    \hfill
    \begin{subfigure}[t]{0.49\textwidth}
        \includegraphics[width=\textwidth]{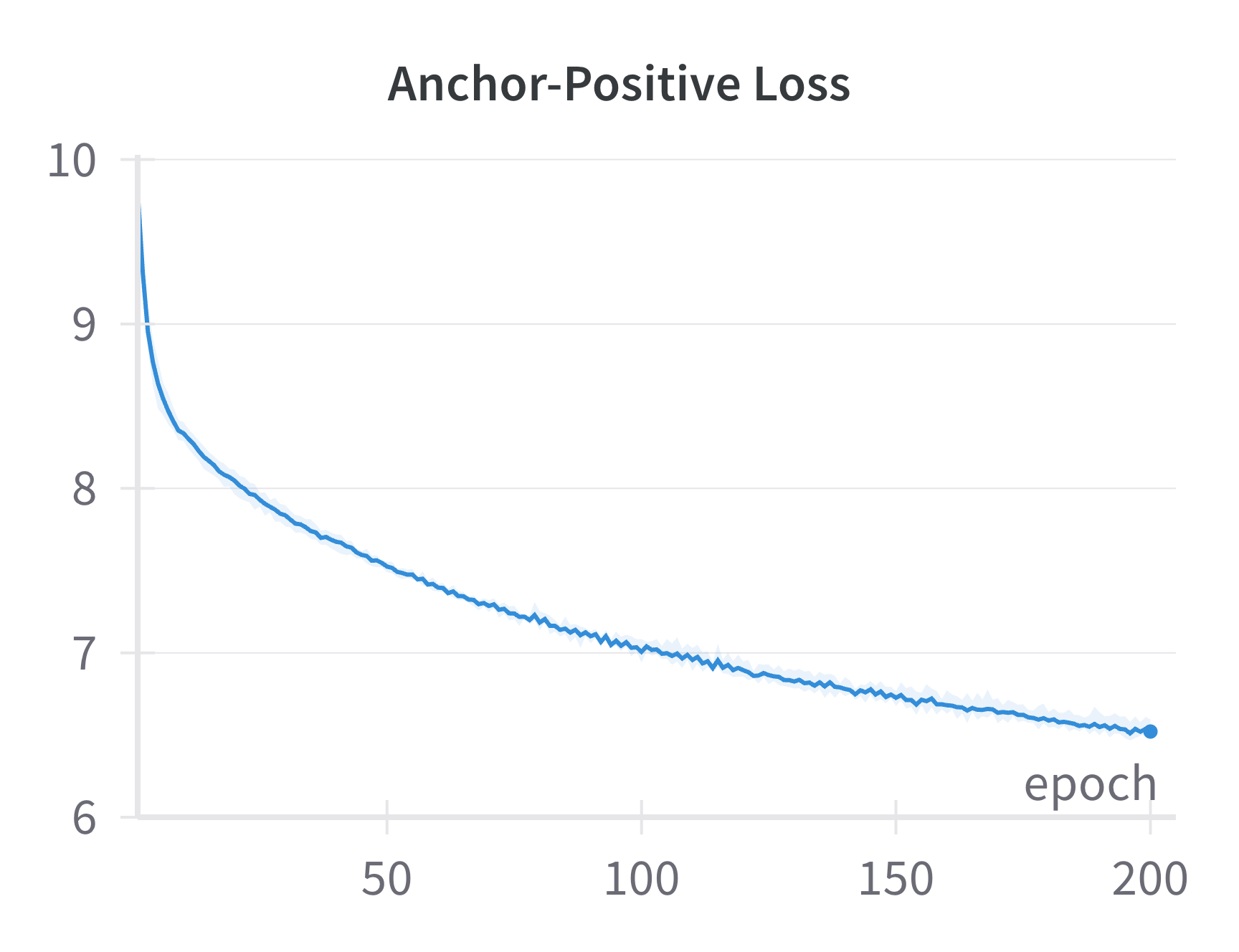}
        \caption{The value of the anchor-positive term.}
    \end{subfigure}
    \begin{subfigure}[t]{0.49\textwidth}
        \includegraphics[width=\textwidth]{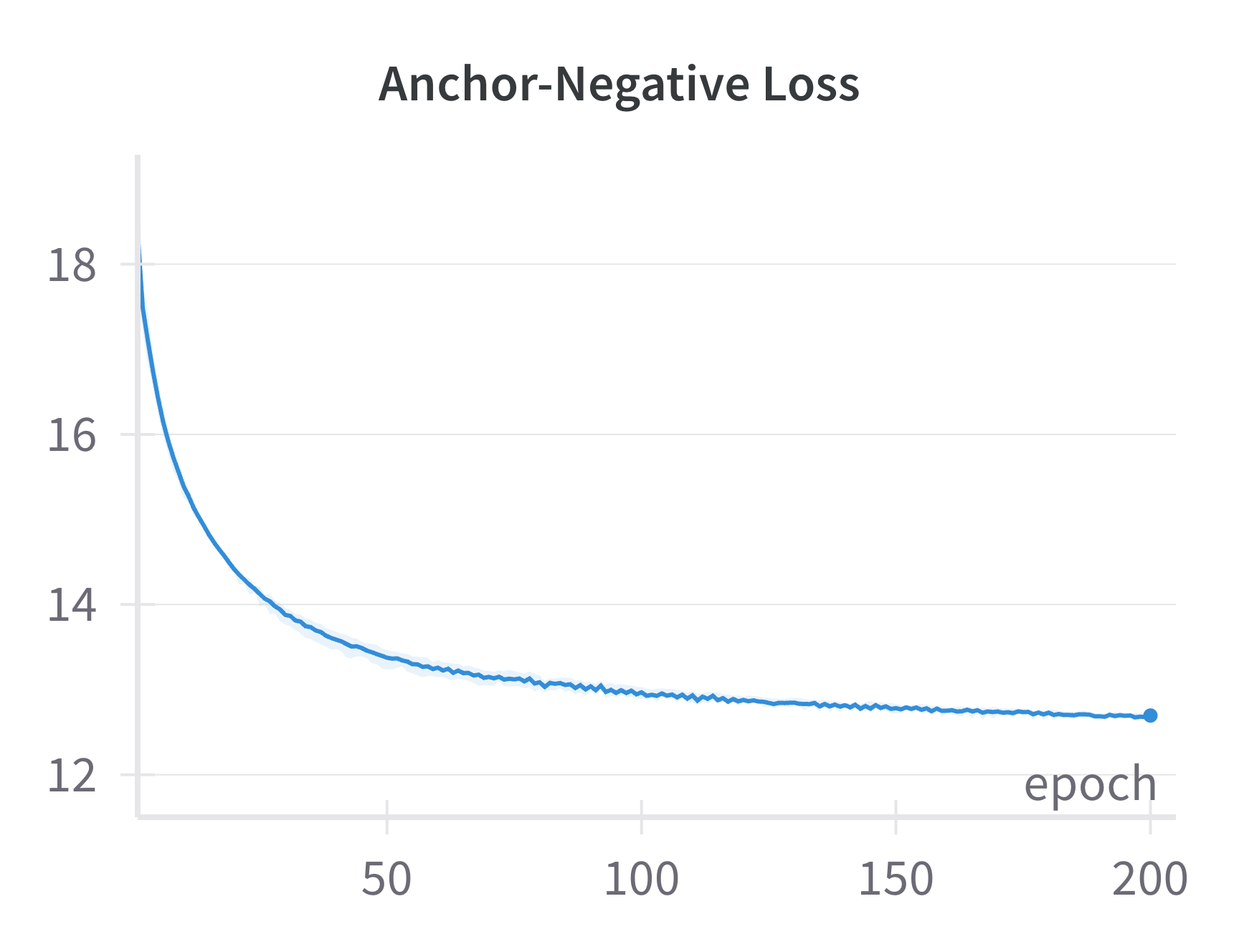}
        \caption{The value of the anchor-negative term.}
    \end{subfigure}
    \hfill
    \begin{subfigure}[t]{0.49\textwidth}
        \includegraphics[width=\textwidth]{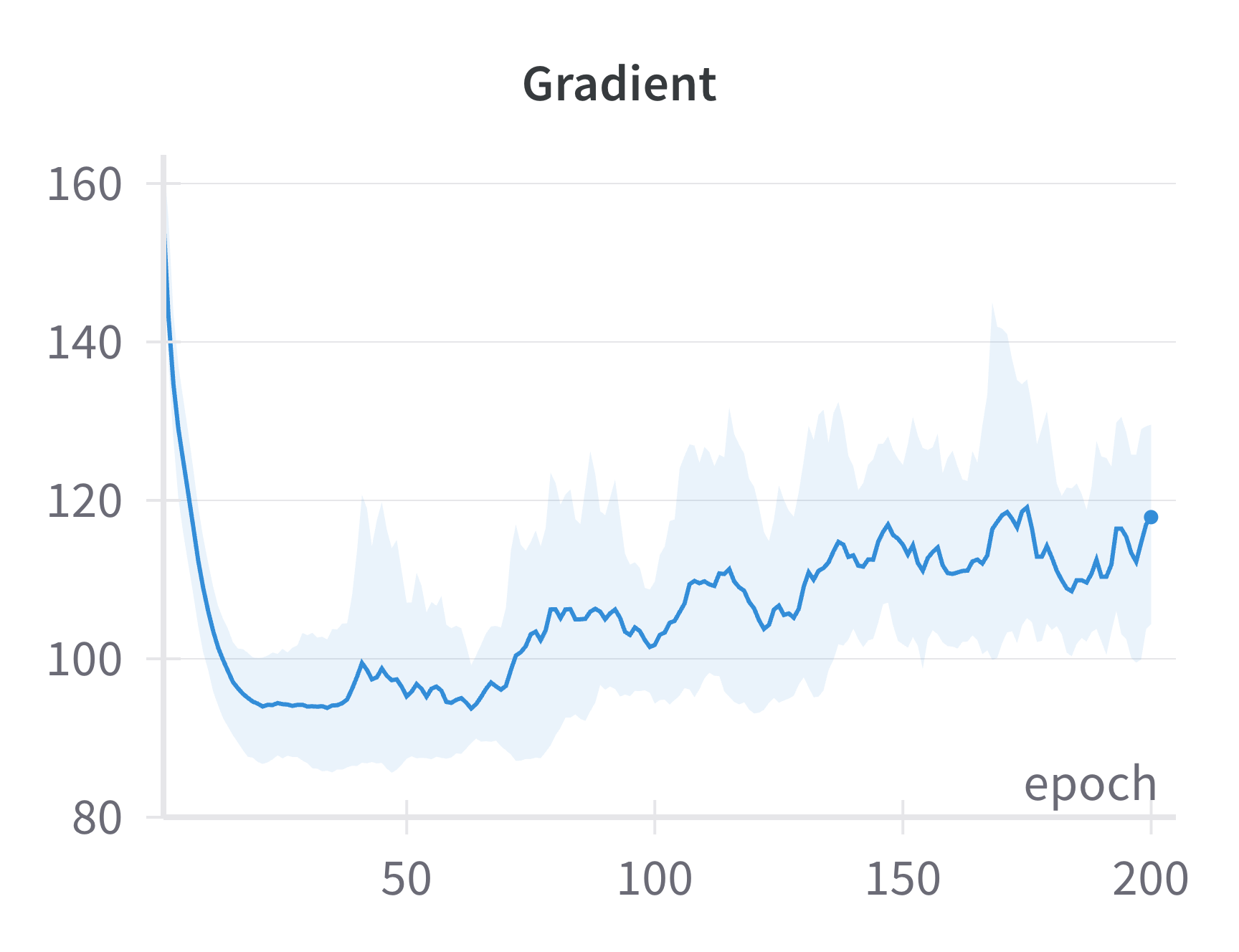}
        \caption{The sum of the gradient of total loss.}
    \end{subfigure}
    \begin{subfigure}[t]{0.49\textwidth}
        \includegraphics[width=\textwidth]{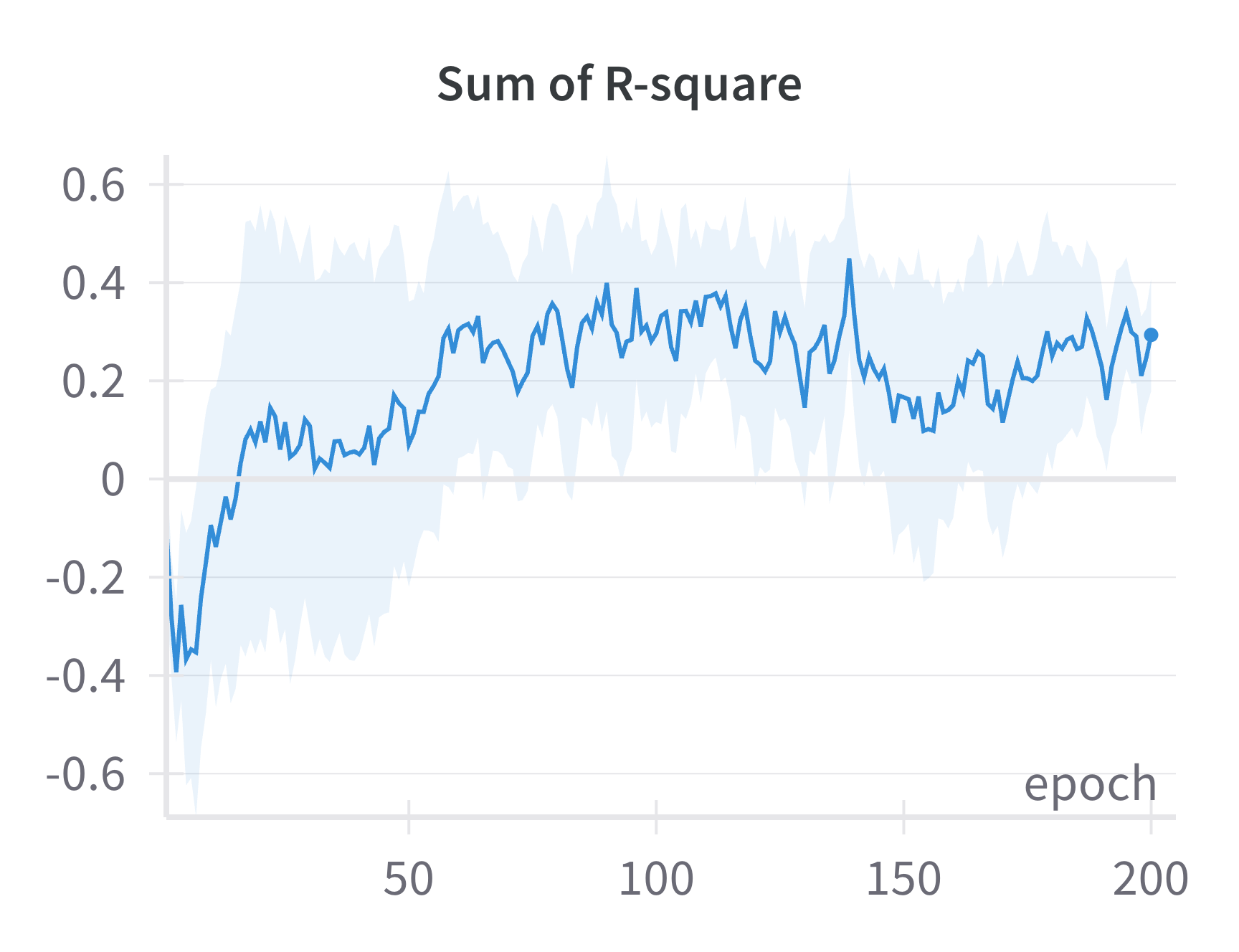}
        \caption{The summation of $r^2$ across the four validation tasks. }
    \end{subfigure}
    \hfill
    \begin{subfigure}[t]{0.49\textwidth}
        \includegraphics[width=\textwidth]{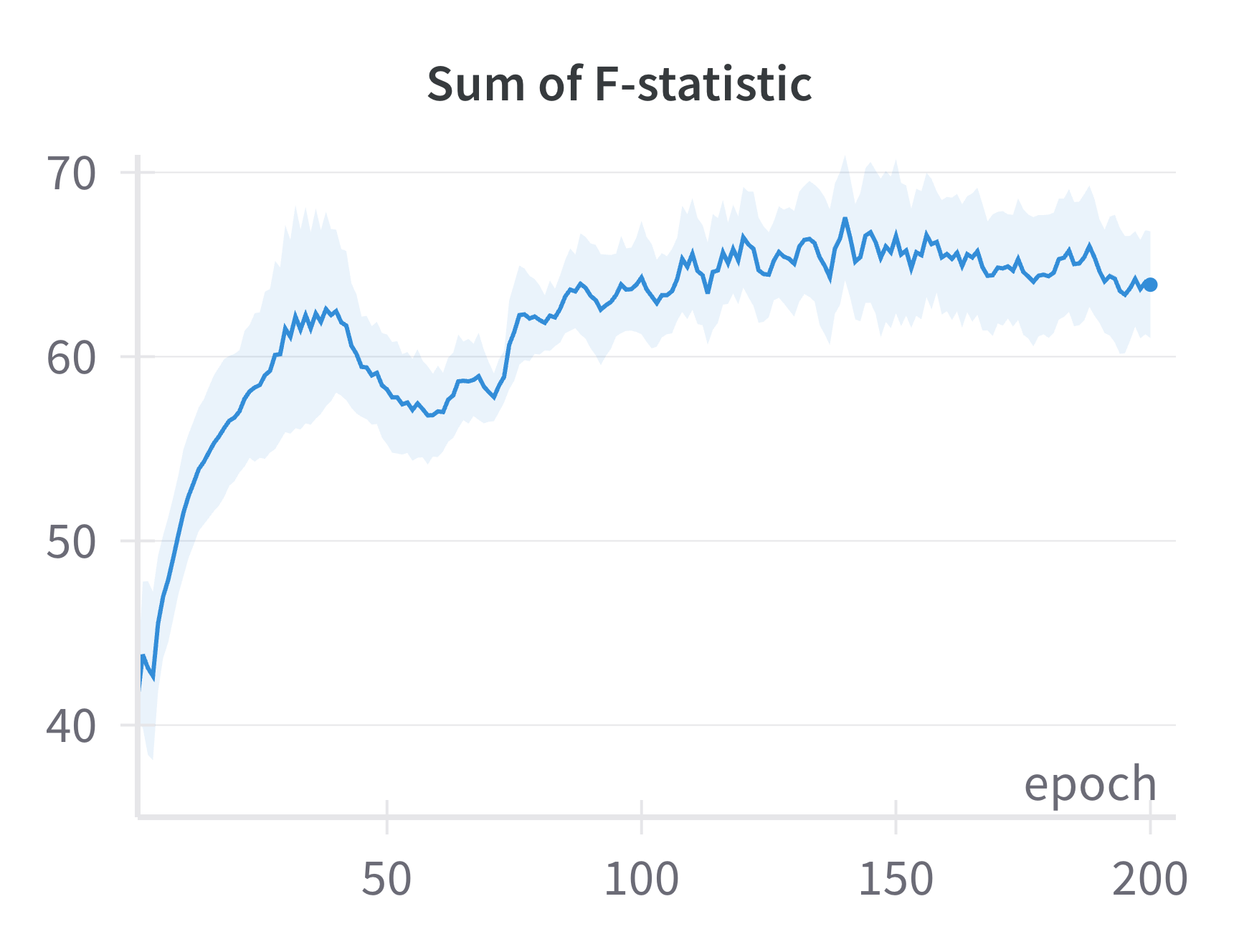}
        \caption{The summation of the Pearson correlation coefficient across the four validation tasks.}
    \end{subfigure}
    \caption{The learning curves of the pre-training process. The mean and range of four curves with distinct random seeds are reported. }
    \label{fig:curves}
\end{figure}

The GIN network was trained with our substrate scope contrastive loss, as described in Equation \ref{eq:loss}. The value of the total loss, anchor-positive term, and anchor-negative term are depicted in Figure \ref{fig:curves}(a-c). The cumulative gradient of the total loss across all trainable parameters is illustrated in Figure \ref{fig:curves}(d). As described in the Method section, under the \textit{Substrate scope contrastive learning} subsection, we constructed four validation tasks to monitor the training process. This monitoring involved evaluating the network's ability to predict reaction yields, Hammett constants, Mulliken charges, and NMR shielding constants. The summation of the coefficient of determination ($r^2$) values across these four validation tasks, as well as the Pearson correlation coefficient between the predicted targets and their most correlated features, are presented in Figure \ref{fig:curves}(e-f). Overall, we can see the $r^2$ and the Pearson correlation coefficient plateau after 50 epochs and oscillate after that. We terminated the training at the 56$^{th}$ epoch, which reached the highest aggregate $r^2$ value.  

We tuned the hyper-parameter of the network architecture and the training details to maximize the summation of $r^2$ across these four validation tasks. The resulting GIN network has 64 hidden channels and 5 layers. An Adam optimizer was used with a momentum of 0.9, an initial learning rate of 0.00005, and a learning rate decay of 0.999. 16 triplets were sampled for each anchor molecule, and a batch size of 4096 was used for training. In Equation \ref{eq:loss}, we used $\gamma$ of 4.023, $\beta$ of 0.612, and $M$ of 1.906.

We show the correlation between the learned embeddings and the Hammett constants in Figures S8–S10. It is important to note that, even for molecules seen during pretraining, the model does not have access to the Hammett constant values, and all metrics reported are for the “test set” in the conventional supervised learning setting. This demonstrates that the model has learned transferable patterns reflecting reactivity through substrate scope contrastive learning.

We also want to highlight that 42 of the 61 molecules with labeled Hammett constants were seen during pretraining. This is because our pretraining dataset includes all reactions involving aryl halides with at least five substrates, which means that most commonly used molecules were already encountered. Consequently, there may be a distribution shift between molecules seen during pretraining and those not seen.

\begin{figure}[H]
    \centering
    \includegraphics[width=0.75\textwidth]{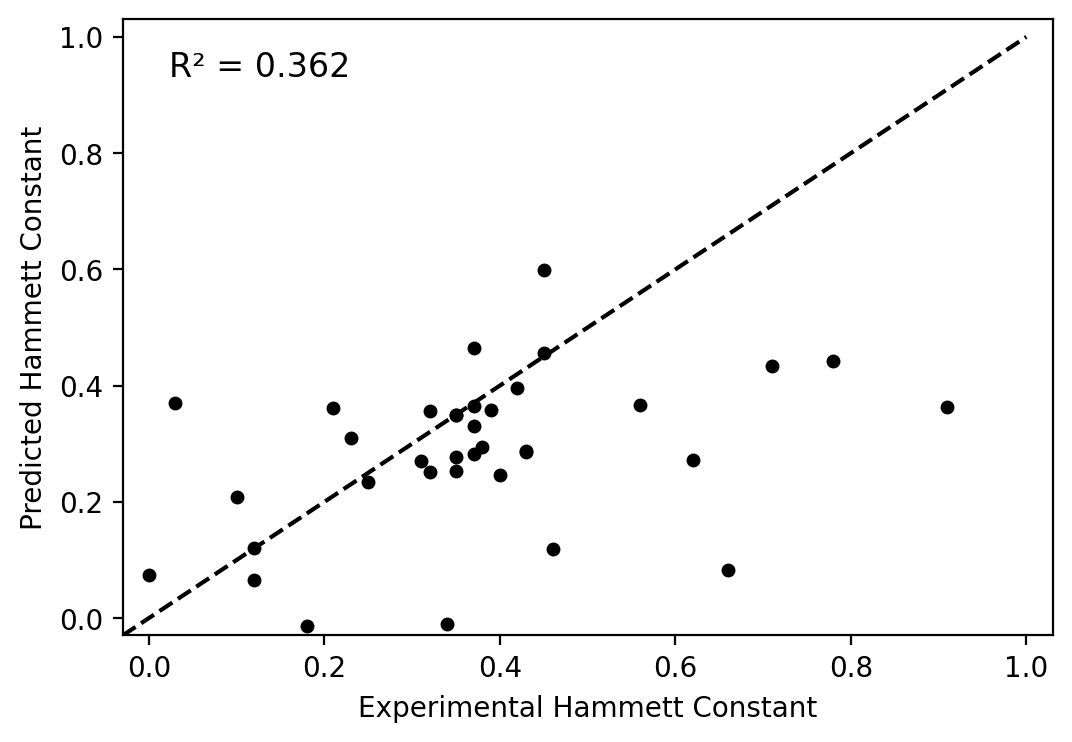}
    \caption{Results of predicting Hammett constants from learned embeddings.}
    \label{fig:hammett_pair}
\end{figure}

\begin{figure}[H]
    \centering
    \includegraphics[width=0.75\textwidth]{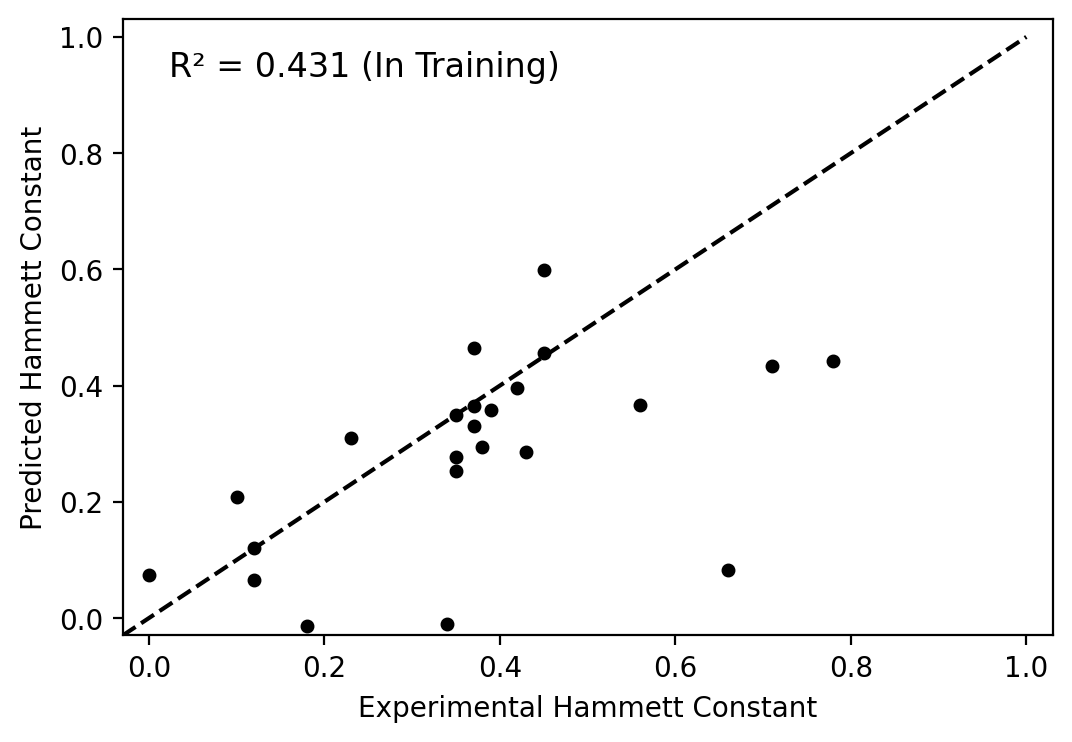}
    \caption{Results of predicting Hammett constants from learned embeddings. This figure shows a subset of data where the aryl halides are seen during pre-training.}
    \label{fig:hammett_pair_in_train}
\end{figure}

\begin{figure}[H]
    \centering
    \includegraphics[width=0.75\textwidth]{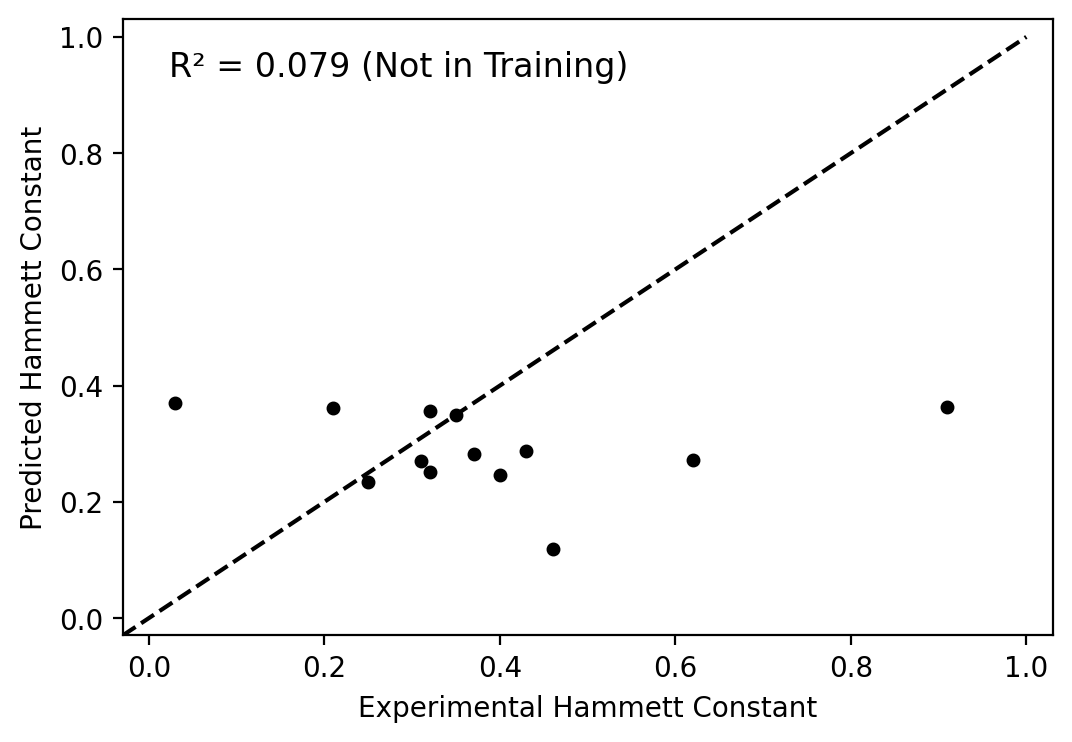}
    \caption{Results of predicting Hammett constants from learned embeddings. This figure shows a subset of data where the aryl halides are not seen during pre-training.}
    \label{fig:hammett_pair_not_in_train}
\end{figure}

\section{Details of reactivity descriptors}

In the following tables, we present descriptions of all reactivity descriptors extracted from DFT-level calculations carried out on the aryl halides.

\begin{figure}[H]
    \centering
    \includegraphics[width=0.8\textwidth]{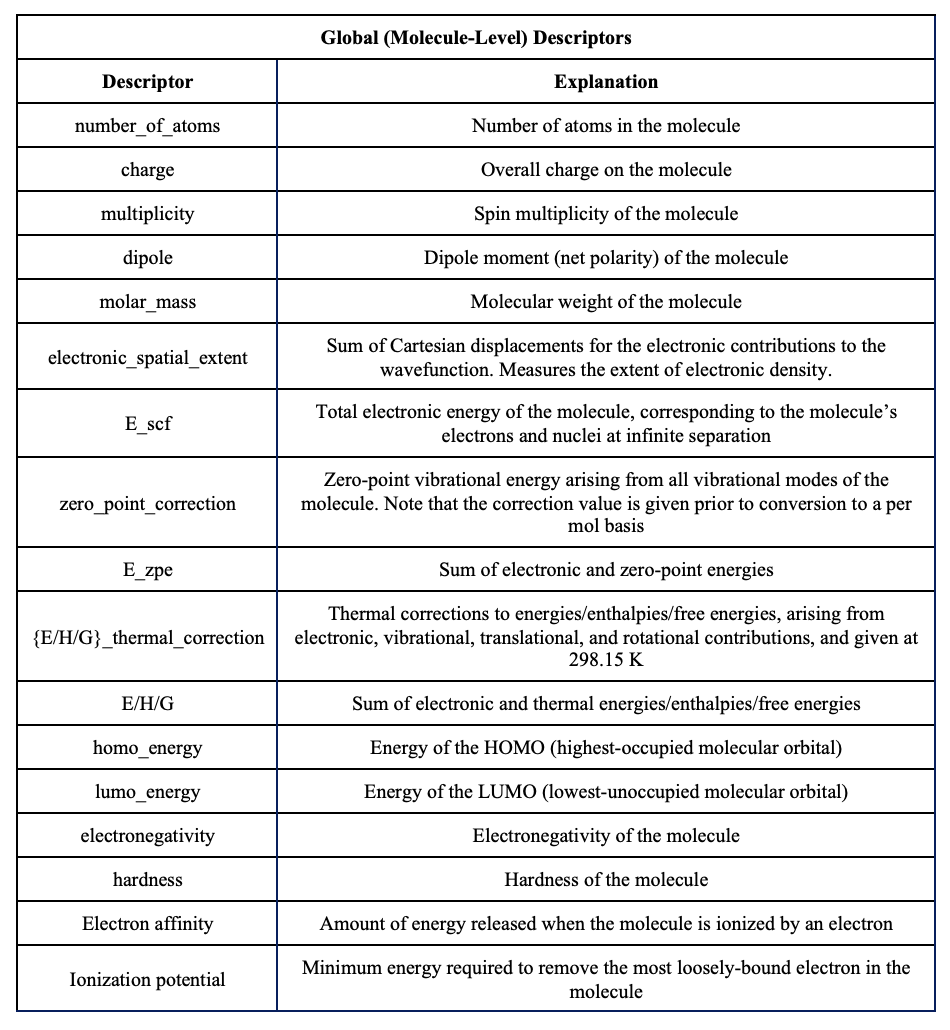}
    \caption{All computed global (molecule-level) descriptors using auto-qchem \cite{zuranski2022auto} and an explanation of each.}
    \label{fig:global_desc}
\end{figure}

\begin{figure}[H]
    \centering
    \includegraphics[width=0.8\textwidth]{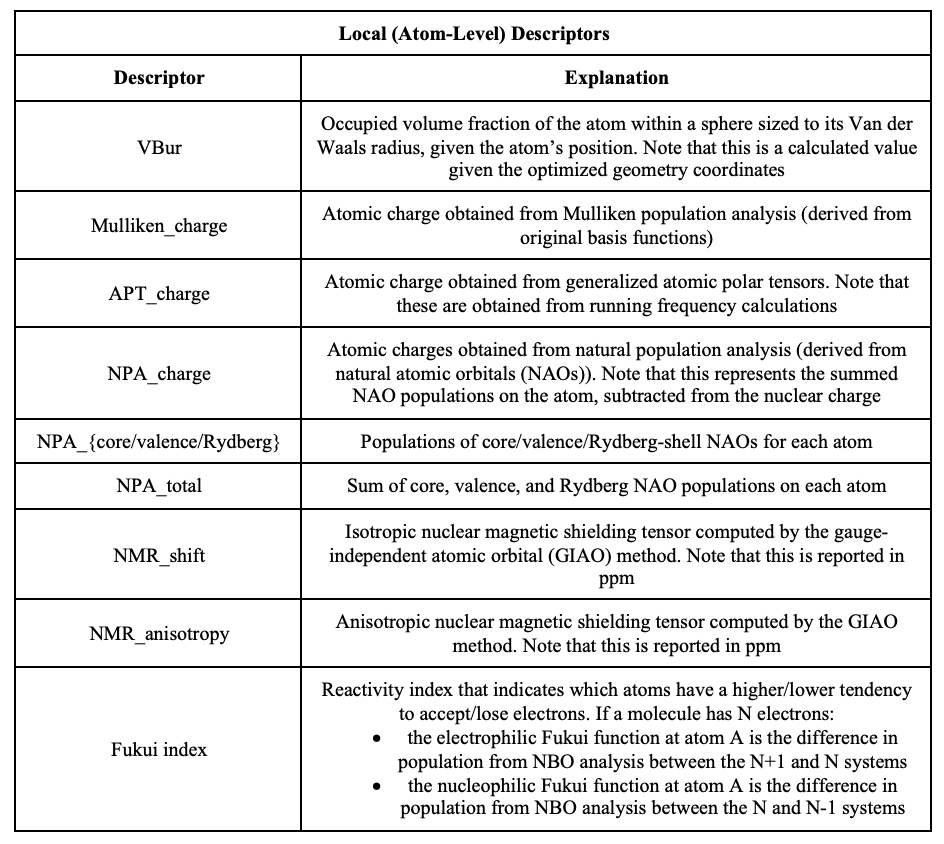}
    \caption{All computed local (atom-level) descriptors using auto-qchem\cite{zuranski2022auto} and an explanation of each.}
    \label{fig:local_desc}
\end{figure}

\section{Additional Results}

\subsection{Qualitative investigation of the learned embeddings}

\begin{figure}[H]
    \centering
    \includegraphics[width=0.95\textwidth]{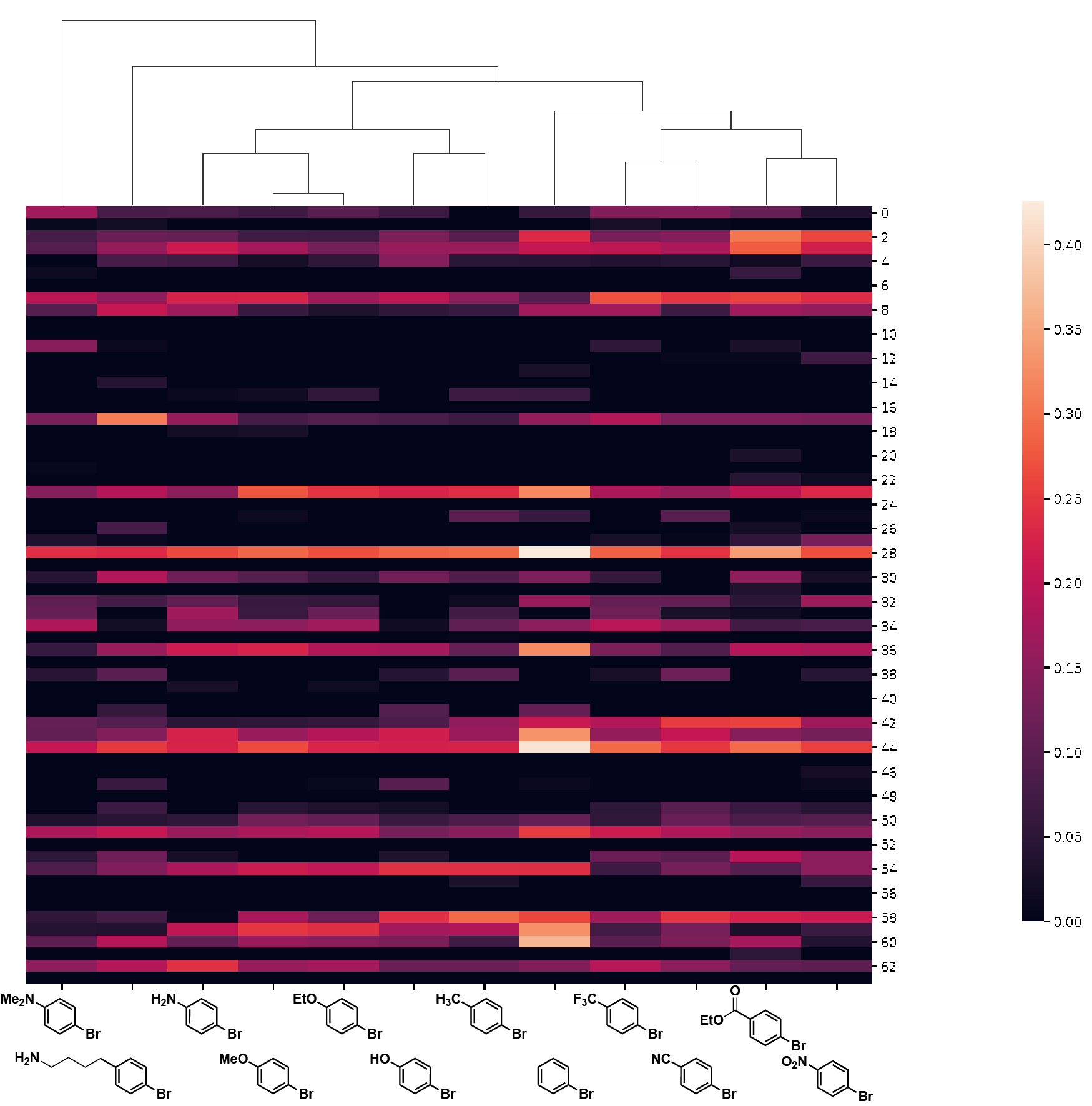}
    \caption{Heatmap of the pair-wise distance between 12 para-substituted aryl bromides, showcasing that the learned embeddings' signal-to-noise (SNR) distances (A) better reflects functional similarity than the conventional Tanimoto distances based on structural fingerprints (B).}
    \label{fig:Fig_bit_viz}
\end{figure}

To offer an intuitive examination of the bit-level details of the embeddings, we illustrate the bit value of the corresponding position in the embeddings with a heatmap and their hierarchical clustering in Figure \ref{fig:Fig_bit_viz}. We analyzed a set of 12 hand-selected para-substituted bromobenzenes, as shown in the figure.

\begin{figure}[t!]
    \centering
    \includegraphics[max width=1.00\textwidth]{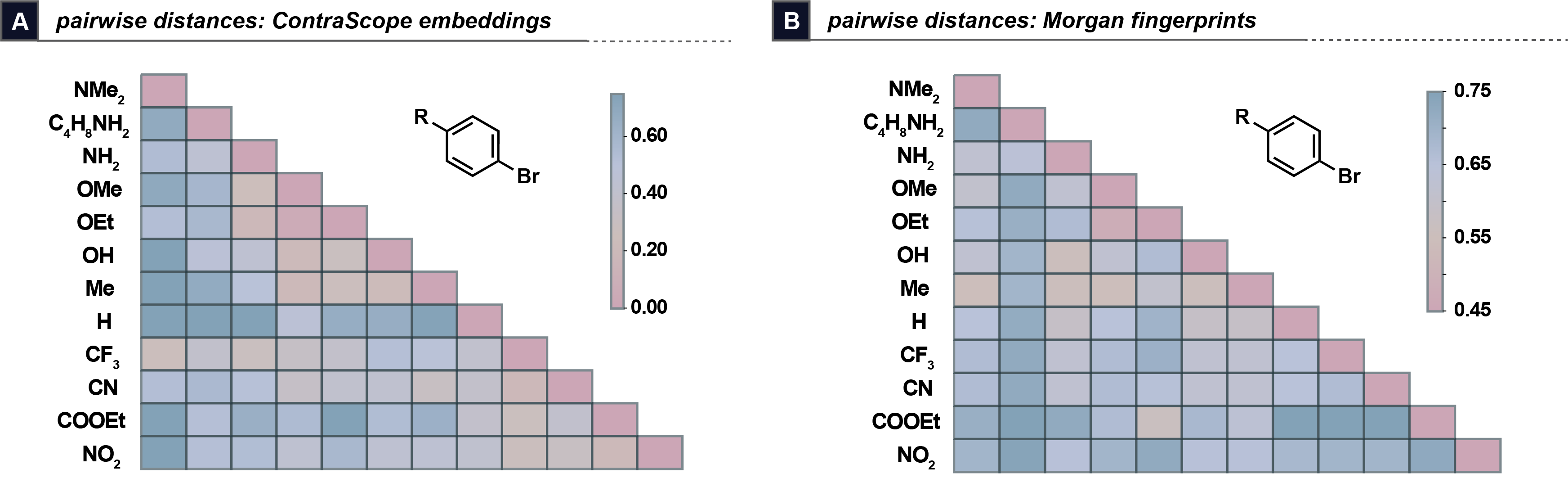}
    \caption{Pairwise distance heatmaps for 12 para-substituted bromobenzenes. (A) Distances calculated using ContraScope embeddings reveal clear clustering of molecules based on functional reactivity, with electron-withdrawing groups showing reduced distances compared to electron-donating groups. (B) Distances derived from Morgan fingerprints using Tanimoto similarity do not exhibit this functional clustering, instead reflecting structural similarity.}
    \label{fig:fig_pair_wise_dist}
\end{figure}

In order to validate whether quantitative distances in the learned embedding space accurately reflect known functional similarity in chemical reactivity, we next encoded the same set of para-substituted bromobenzenes, then calculated and visualized their pairwise distances (Figure \ref{fig:fig_pair_wise_dist}). The heatmap reveals that molecules with electron-withdrawing groups, such as nitrile, ester, and trifluoromethyl, demonstrate notably reduced distances compared to those with electron-donating groups like methoxy, hydroxyl, ether, and amine. This discernible pattern is not mirrored in the Tanimoto distances\cite{chen2002performance,bajusz2015tanimoto} (refer to Figure \ref{fig:fig_pair_wise_dist}), where distances are measured based on shared structural features in Morgan fingerprints. This comparison emphasizes that distances in our learned embedding space more effectively capture functional similarities in chemical reactivity than standard measures of structural similarity.

\subsection{Visualization of learned aryl halide chemical spaces}

\begin{figure}[H]
     \centering
    \begin{subfigure}[t]{\textwidth}
        \includegraphics[width=\textwidth]{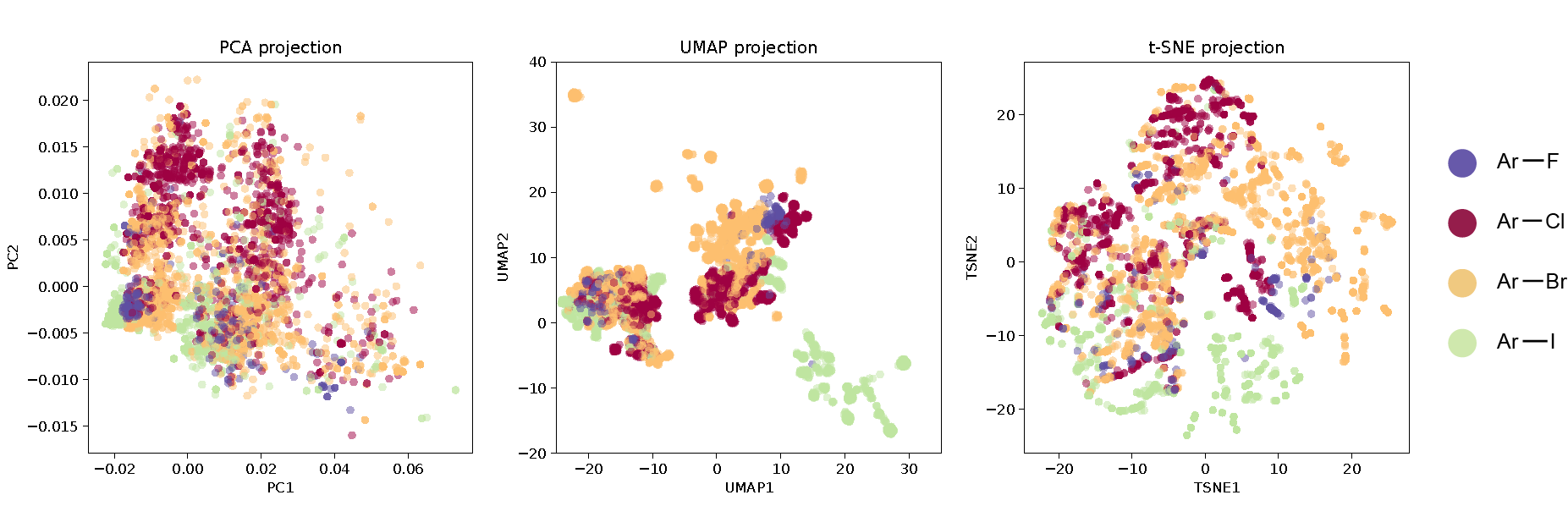}
        \caption{The projection of embeddings before training the GIN using substrate scope groupings.}
        \end{subfigure}
    \hfill
     \begin{subfigure}[t]{\textwidth}
        \includegraphics[width=\textwidth]{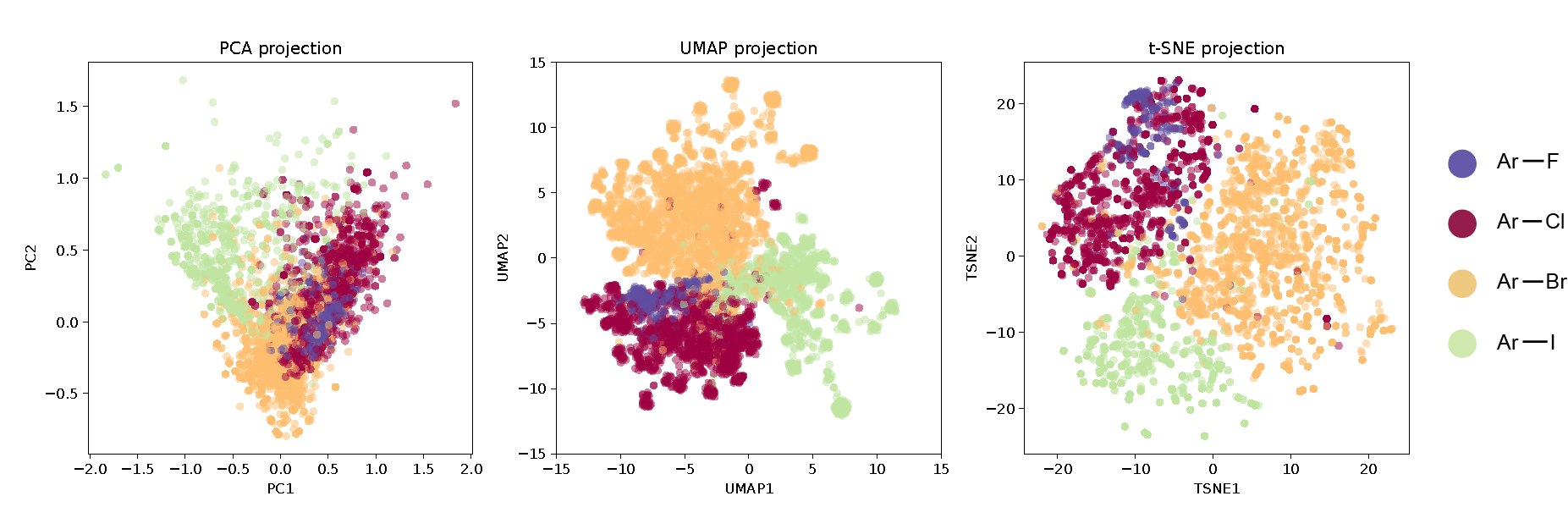}
        \caption{The projection of embeddings after training the GIN using substrate scope groupings.}
        \end{subfigure}
    \caption{The projection of embeddings on a random sampled subset of the substrate scope data. Points are colored by the class of halides.}
    \label{fig:bef_aft}
\end{figure}

To offer insights into the evolution of the embeddings during training, we present the comparative analysis of our embeddings using PCA, t-SNE, and UMAP projections, both pre- and post-training, in Figure \ref{fig:bef_aft}. This analysis utilizes the identical set of molecules featured in Figure \ref{fig:visualization}.

\subsection{Correlation with conventional reactivity indicators}

Below, we show the analysis of SVM models’ regression performance as a function of dataset size on descriptors other than shown in Figure \ref{fig:correlation}C.

\begin{figure}[H]
    \centering
    \includegraphics[width=0.75\textwidth]{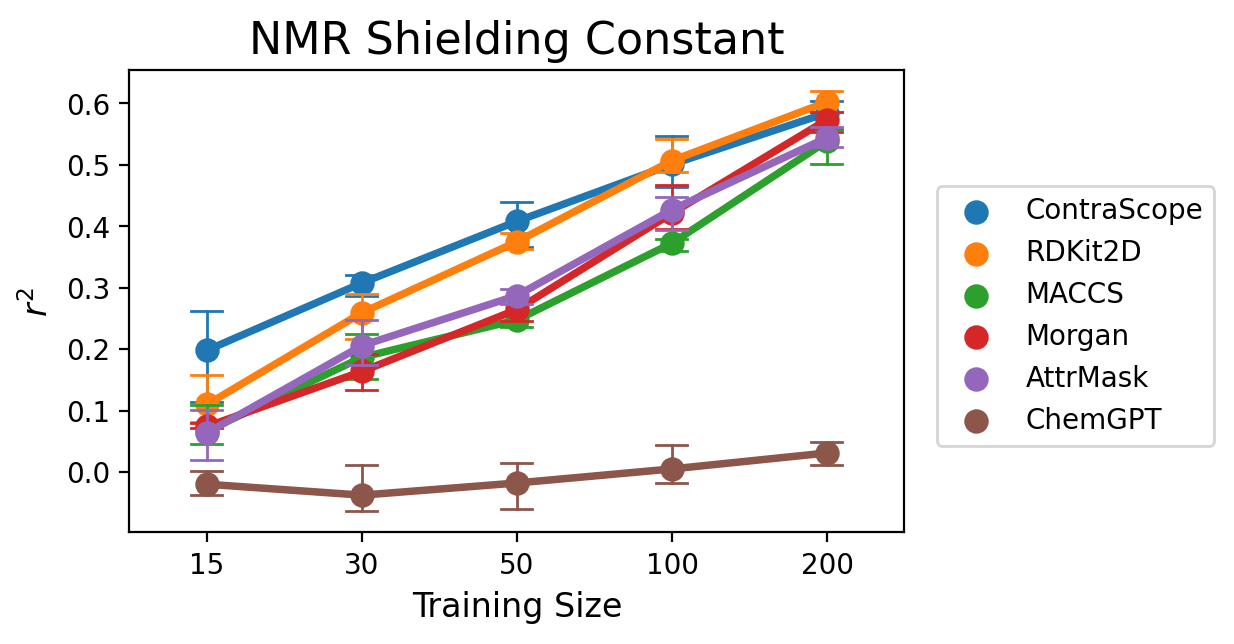}
    \caption{Analysis of SVM regression performance under different dataset size for predicting NMR shielding constants.}
    \label{fig:curve_NMR}
\end{figure}

\begin{figure}[H]
    \centering
    \includegraphics[width=0.75\textwidth]{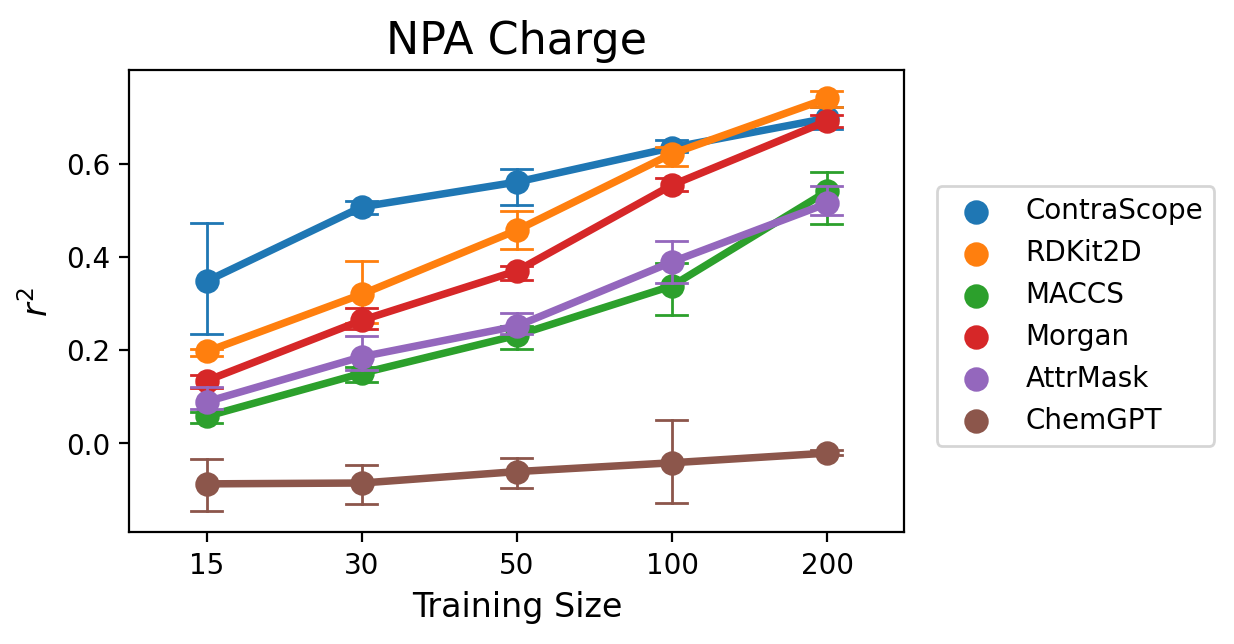}
    \caption{Analysis of SVM regression performance under different dataset size for predicting NPA charges.}
    \label{fig:curve_NPACharge}
\end{figure}

\begin{figure}[H]
    \centering
    \includegraphics[width=0.75\textwidth]{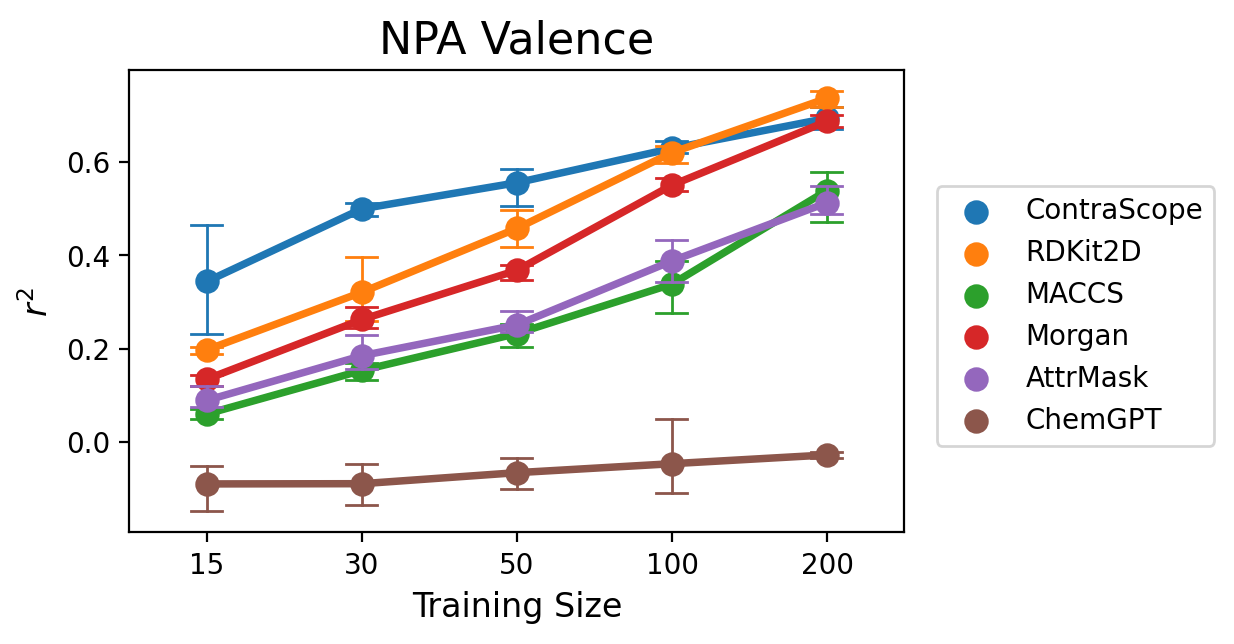}
    \caption{Analysis of SVM regression performance under different dataset size for predicting NPA valences. }
    \label{fig:curve_NPAValence}
\end{figure}

Below, we show the t-SNE projection visualizations of the learned embeddings of the 762 aryl halides under the same setting of Figure \ref{fig:visualization}A, colored by traditional reactivity indicator values.

\begin{figure}[H]
     \centering
    \begin{subfigure}[t]{0.49\textwidth}
        \includegraphics[width=\textwidth]{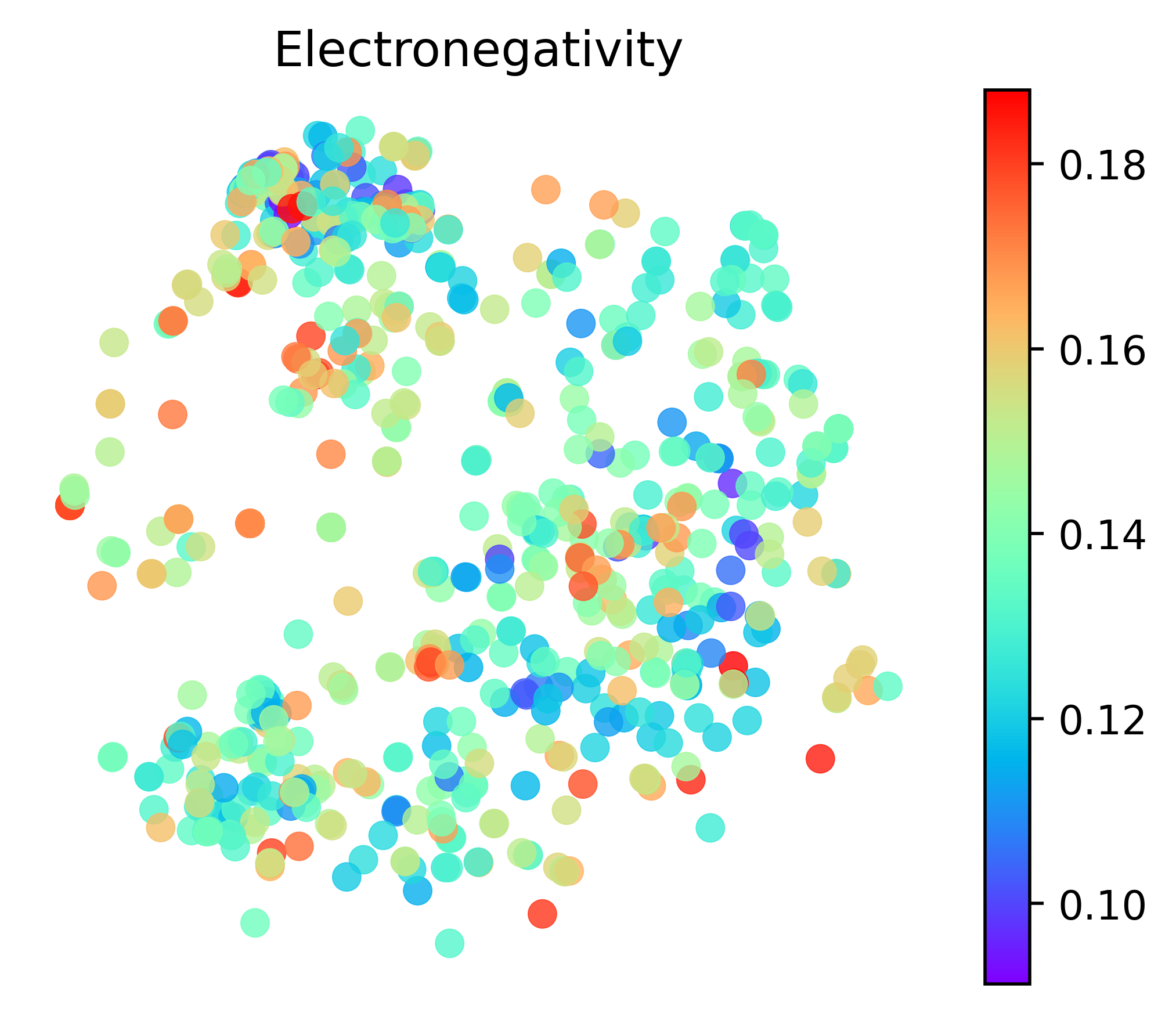}
    \end{subfigure}
    \hfill
    \begin{subfigure}[t]{0.49\textwidth}
        \includegraphics[width=\textwidth]{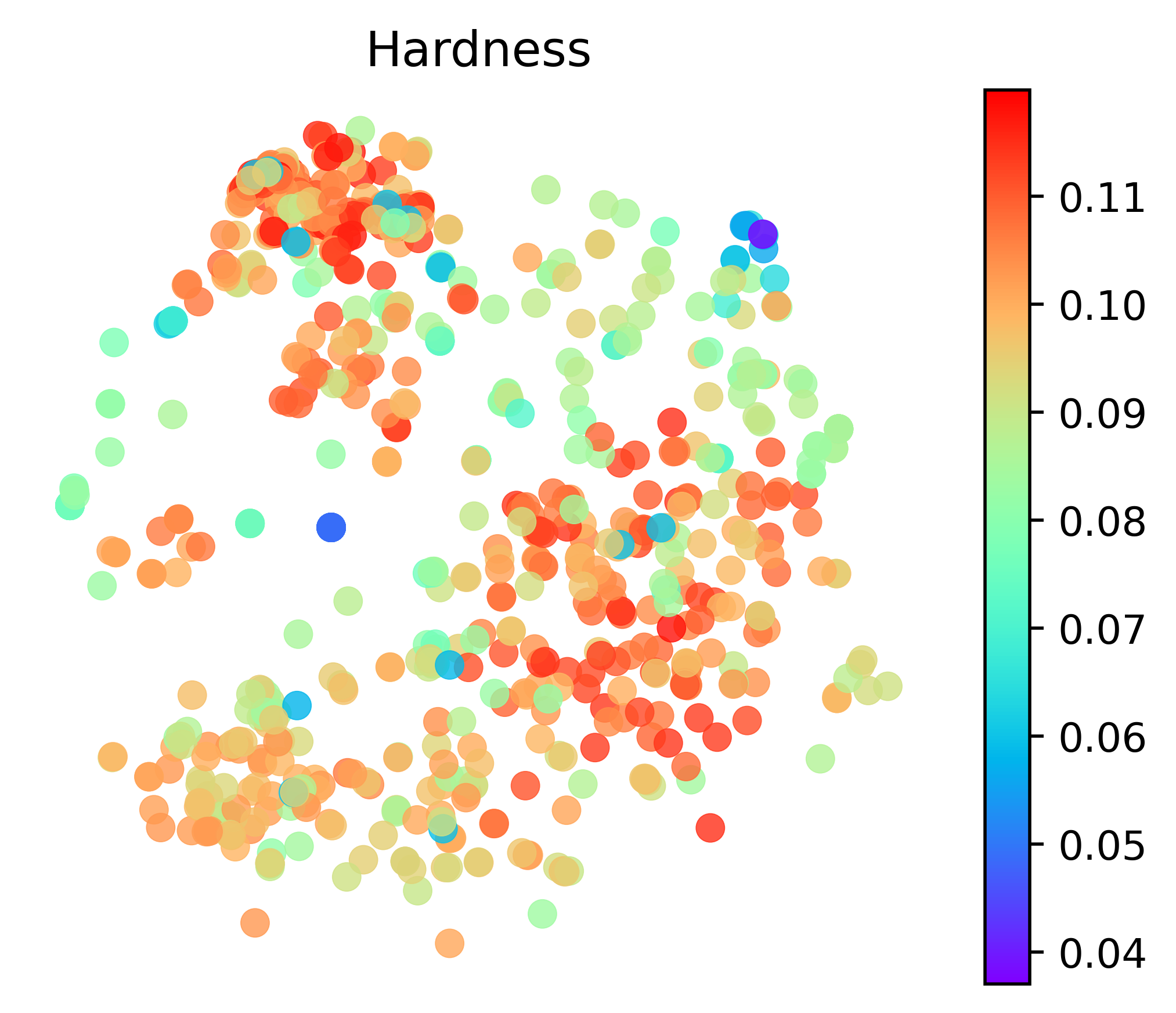}
    \end{subfigure}
    \begin{subfigure}[t]{0.49\textwidth}
        \includegraphics[width=\textwidth]{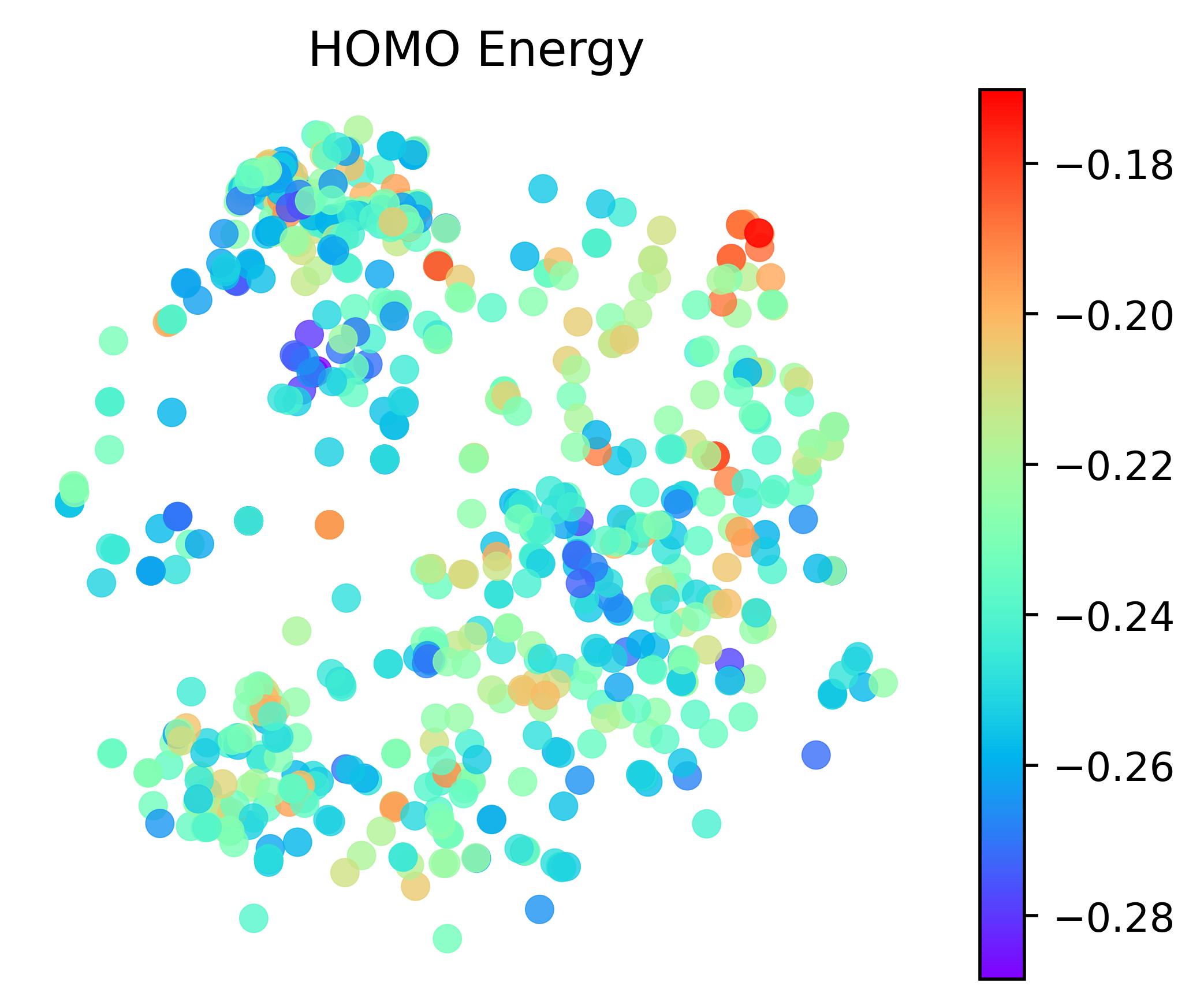}
    \end{subfigure}
    \hfill
    \begin{subfigure}[t]{0.49\textwidth}
        \includegraphics[width=\textwidth]{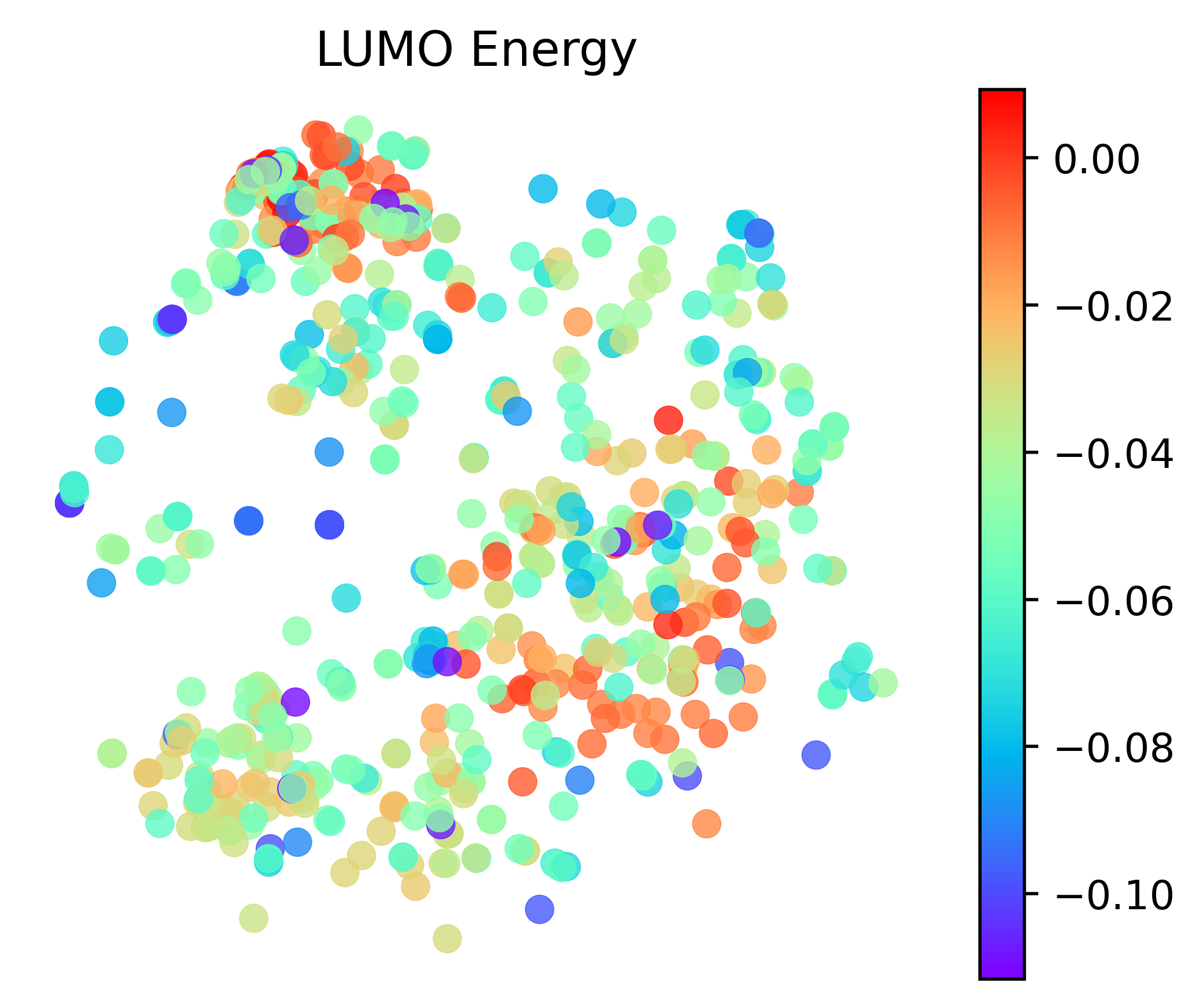}
    \end{subfigure}
    \begin{subfigure}[t]{0.49\textwidth}
        \includegraphics[width=\textwidth]{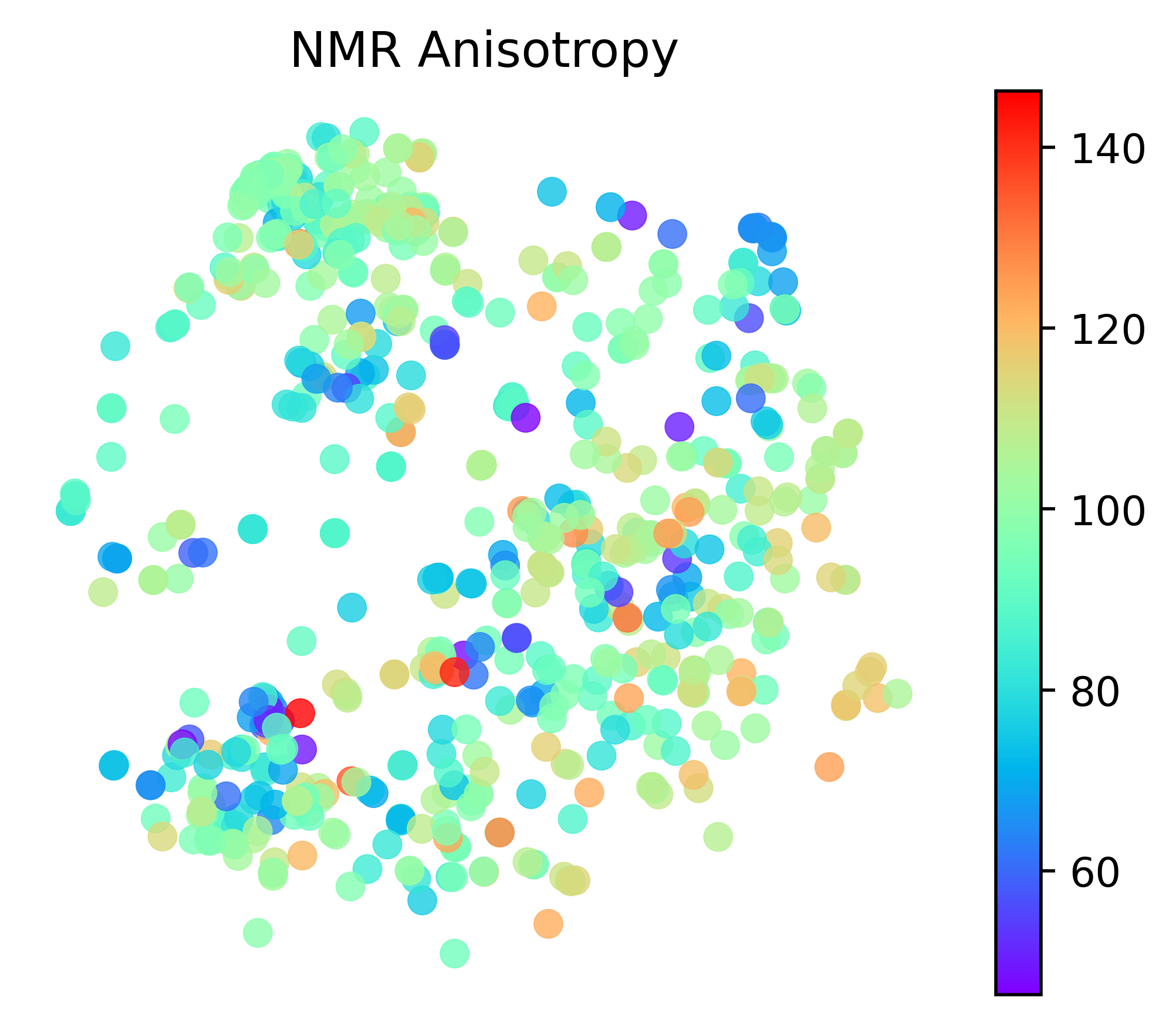}
    \end{subfigure}
    \hfill
    \begin{subfigure}[t]{0.49\textwidth}
        \includegraphics[width=\textwidth]{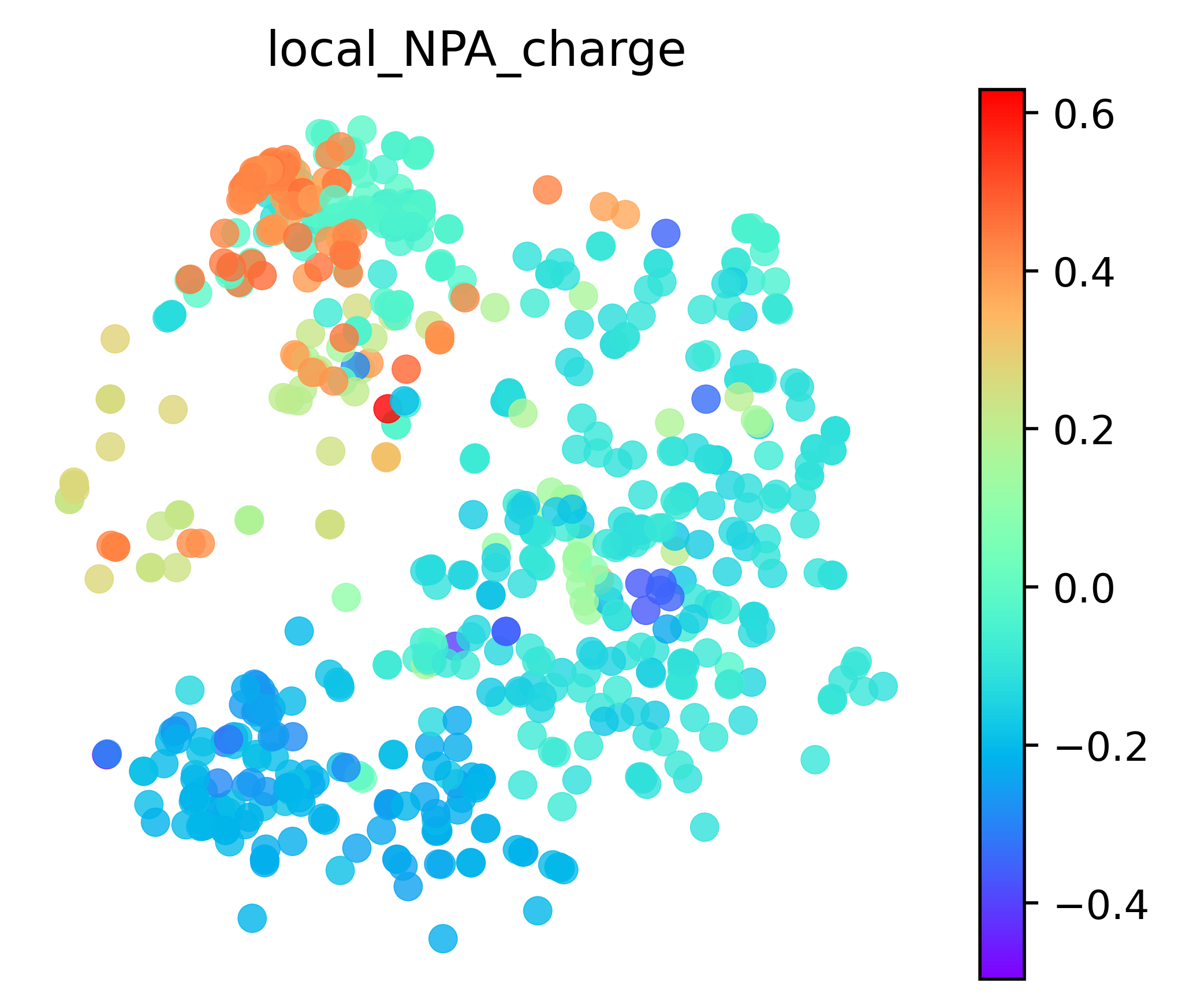}
    \end{subfigure}
    \caption{The t-SNE projection of learned embeddings with each point colored with various reactivity descriptors. }
    \label{fig:tsne}
\end{figure}

\begin{figure}[H]
     \centering
    \begin{subfigure}[t]{0.49\textwidth}
        \includegraphics[width=\textwidth]{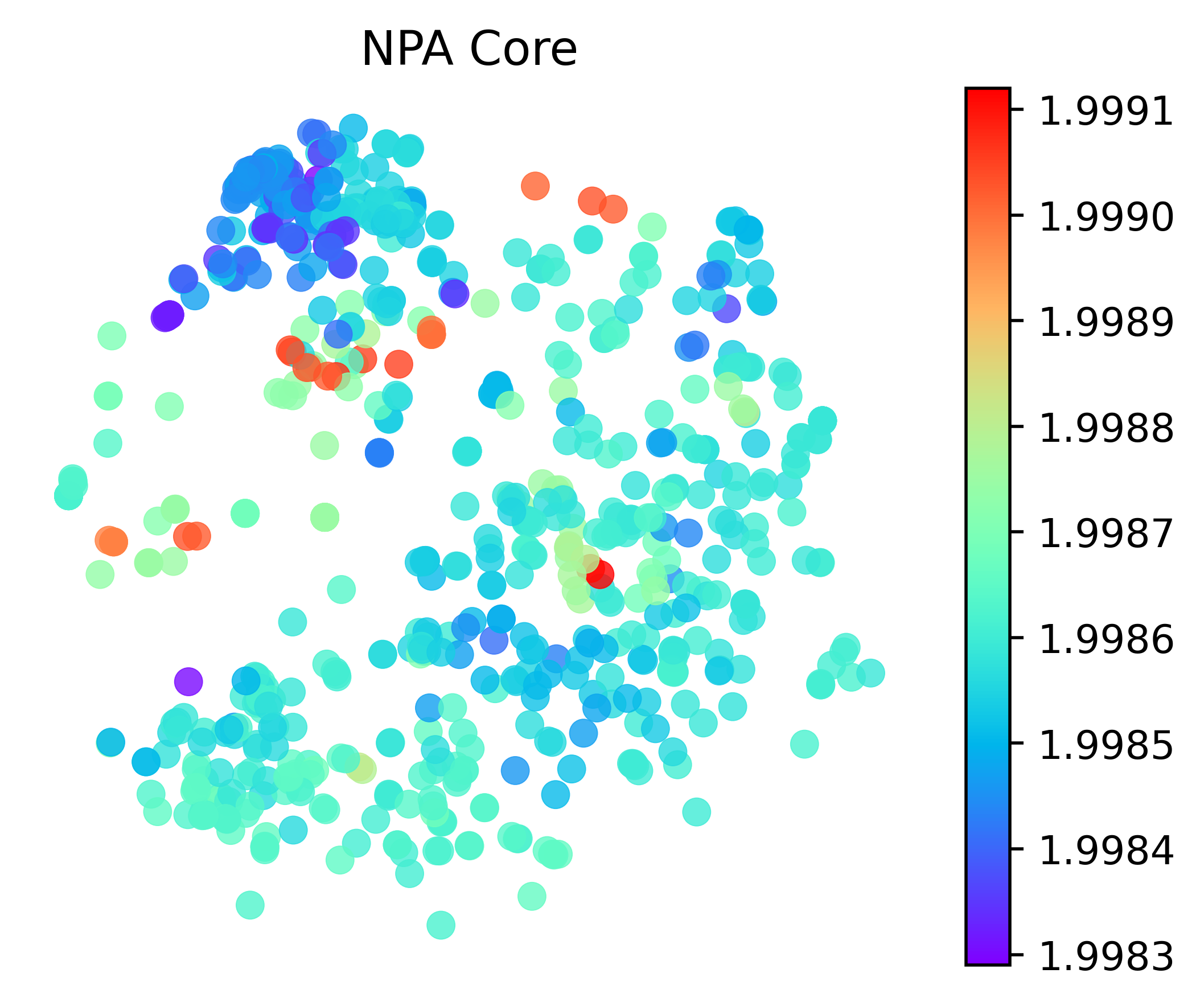}
    \end{subfigure}
    \hfill
    \begin{subfigure}[t]{0.49\textwidth}
        \includegraphics[width=\textwidth]{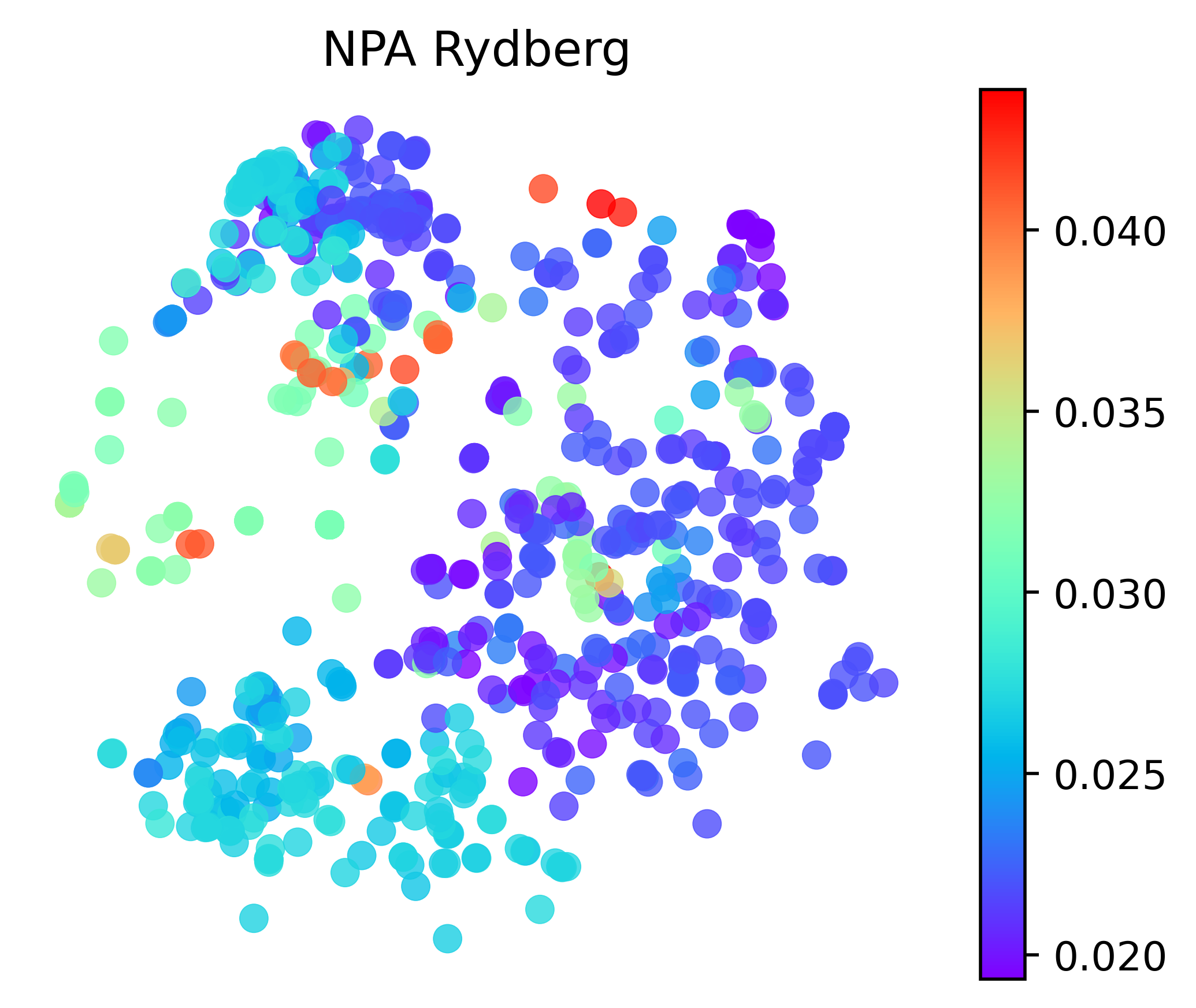}
    \end{subfigure}
        \includegraphics[width=0.49\textwidth]{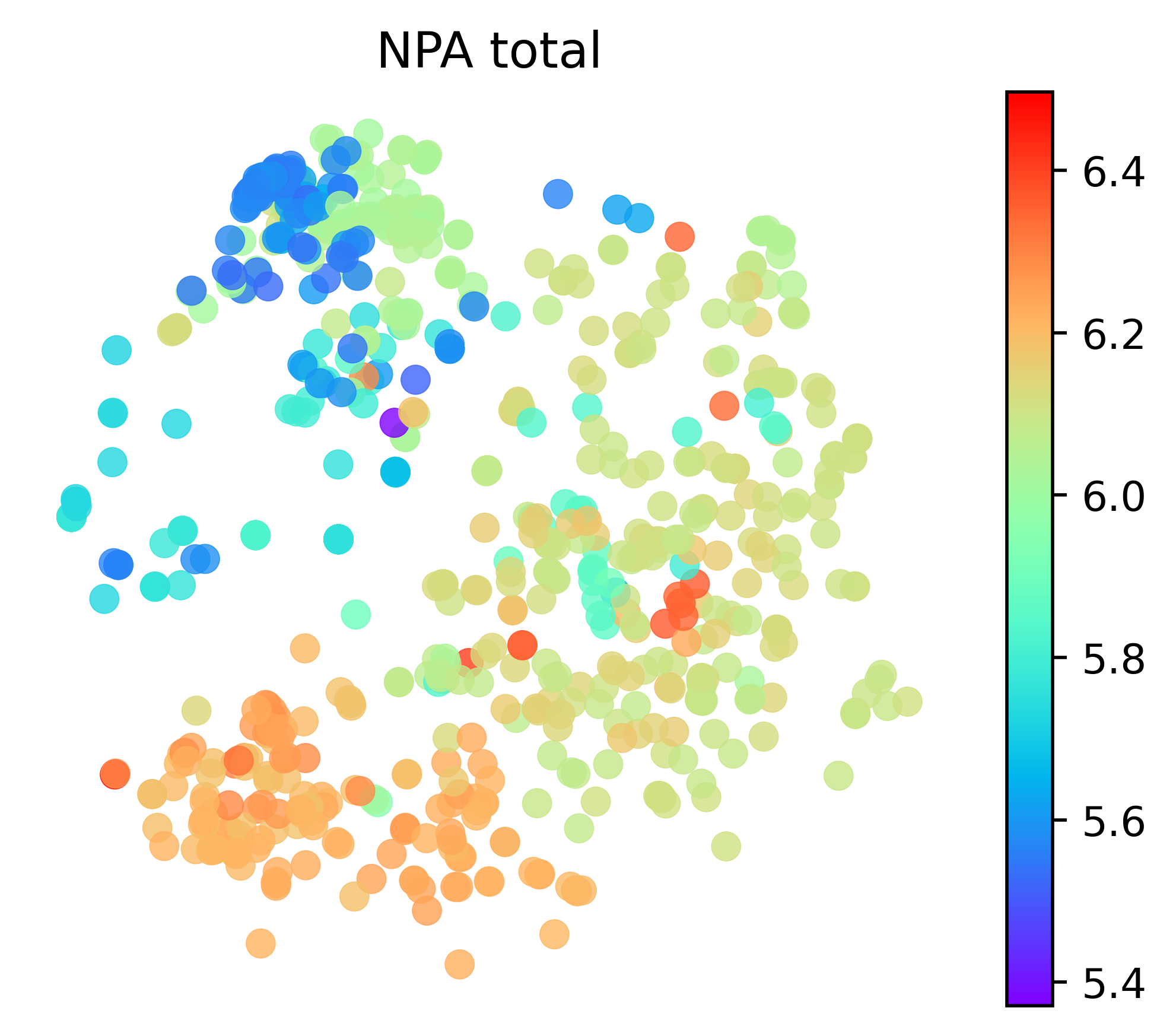}
    \caption{The t-SNE projection of learned embeddings with each point colored with various reactivity descriptors. (Continued) }
    \label{fig:tsne2}
\end{figure}

\subsection{Regression performance on the training substrate scopes}

We tried to use the learned embedding as feature to predict reaction yields in the training data. Each substrate scope was treated as a single regression task and leave-one-out cross validation $r^2$ for each scope is shown. The methods compared include ContraScope combined with k-Nearest Neighbors (kNN), RDKit2D with kNN, and RDKit2D with Random Forest (RF). The distributions highlight the variability and challenges encountered in the predictive modeling of chemical yields, with all approaches showing a wide distribution of $r^2$ values, including negative values indicative of a failure to capture any trend. 

\begin{figure}[H]
    \centering
    \includegraphics[width=0.5\textwidth]{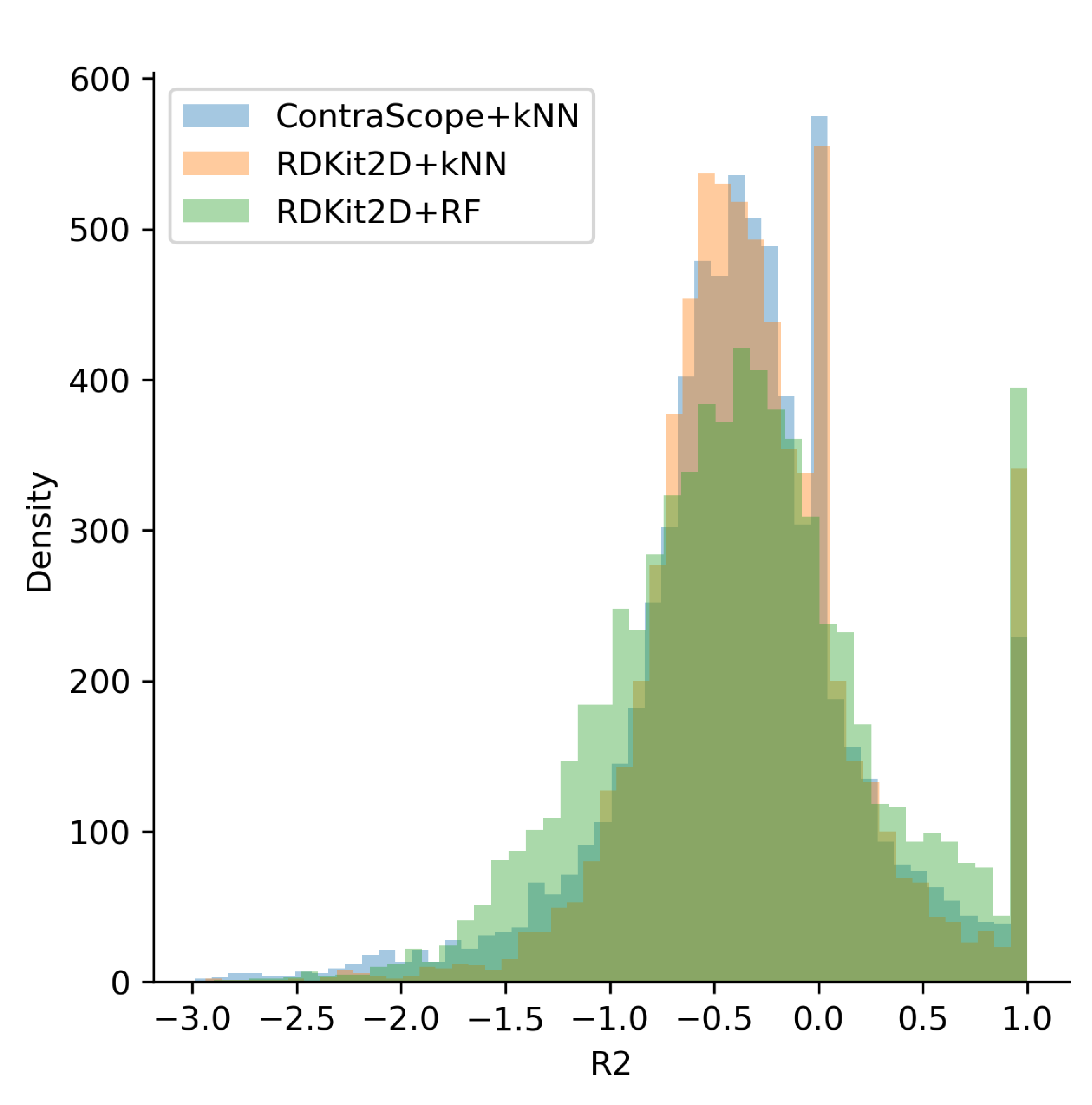}
    \caption{Distribution of the $r^2$ values for yield prediction across all substrate scopes using a leave-one-out cross-validation to evaluate each. None of the representations tested provides a promising degree of predictivity.}
    \label{fig:cas_r2_dist}
\end{figure}

\subsection{Investigation on chemistry informers}

\begin{figure}[H]
    \centering
    \includegraphics[width=\textwidth]{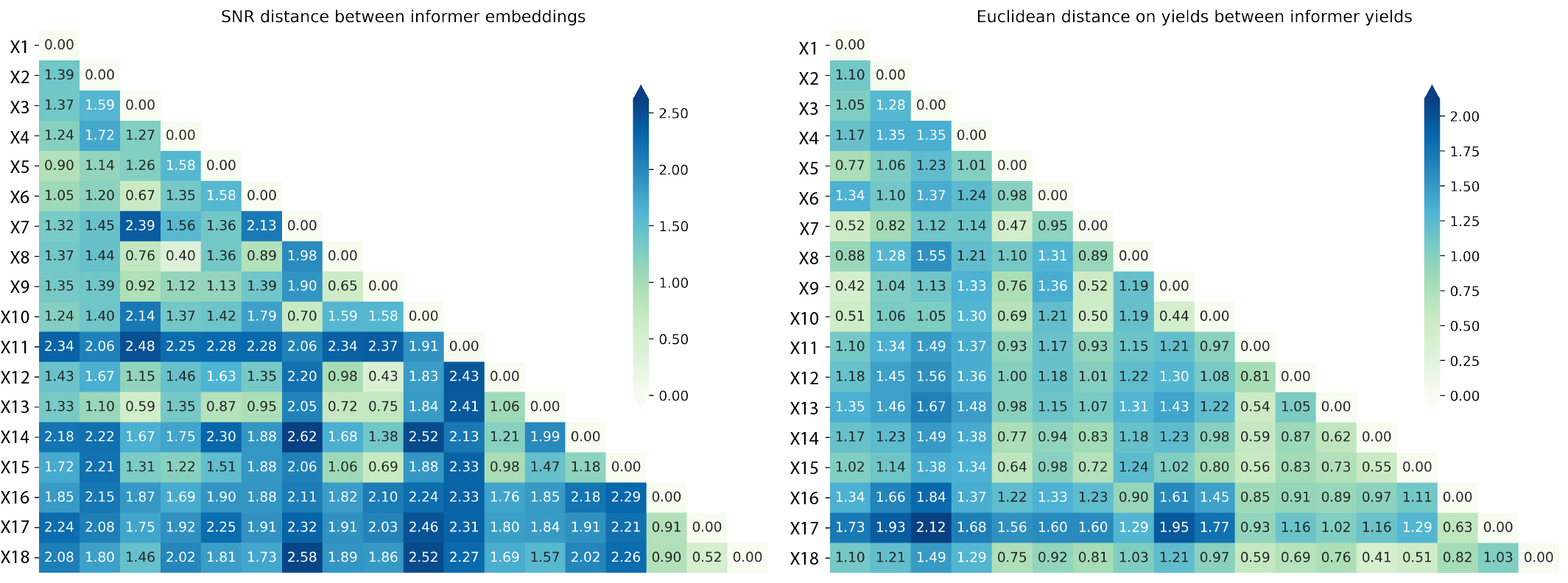}
    \caption{Comparison between the pair-wise SNR distance between the ContraScope embeddings (left), and the pair-wise Euclidean distance between the reported yield vectors (right). Both pair-wise distance matrices are visualized in heatmaps.}
    \label{fig:Fig_informer}
\end{figure}

In the study by Kutchukian et al. (2016)\cite{kutchukian2016chemistry}, a library of 18 aryl halides was utilized to explore their reactivity in the Buchwald-Hartwig reaction under 18 distinct conditions. We encoded these aryl halides from the specified aryl halide informer library. Subsequently, our analysis involved a comparative assessment of the pair-wise signal-to-noise ratio (SNR) distances derived from the ContraScope embeddings against the pair-wise Euclidean distances computed from the vectors of reported yields. This comparison is graphically represented using heatmaps in Figure \ref{fig:Fig_informer}. However, it is noteworthy that this analysis did not reveal congruent patterns in the left and right panels of the figure.

\subsection{Modified training strategies}
\label{sec:negative_sampling}

To address potential biases in our negative sampling strategy, we explored hard negative sampling approaches. First, we focused on substrates with the same halogen type when generating negative samples, treating substrates excluded from the published scope as negative examples only if their halogen matched that of the substrates selected as the anchor. This adjustment aimed to mitigate differences in halogen reactivity, ensuring that the generated negative examples reflected comparable chemical contexts and reducing confounding effects. Additionally, we narrowed our focus to Pd-catalyzed reactions, based on the NameRXN reaction category, to eliminate noise arising from varying reactivities required for different types of reactions. Despite these efforts, neither strategy yielded meaningful improvements.

\begin{figure}[H]
     \centering
    \begin{subfigure}[t]{0.65\textwidth}
        \includegraphics[width=\textwidth]{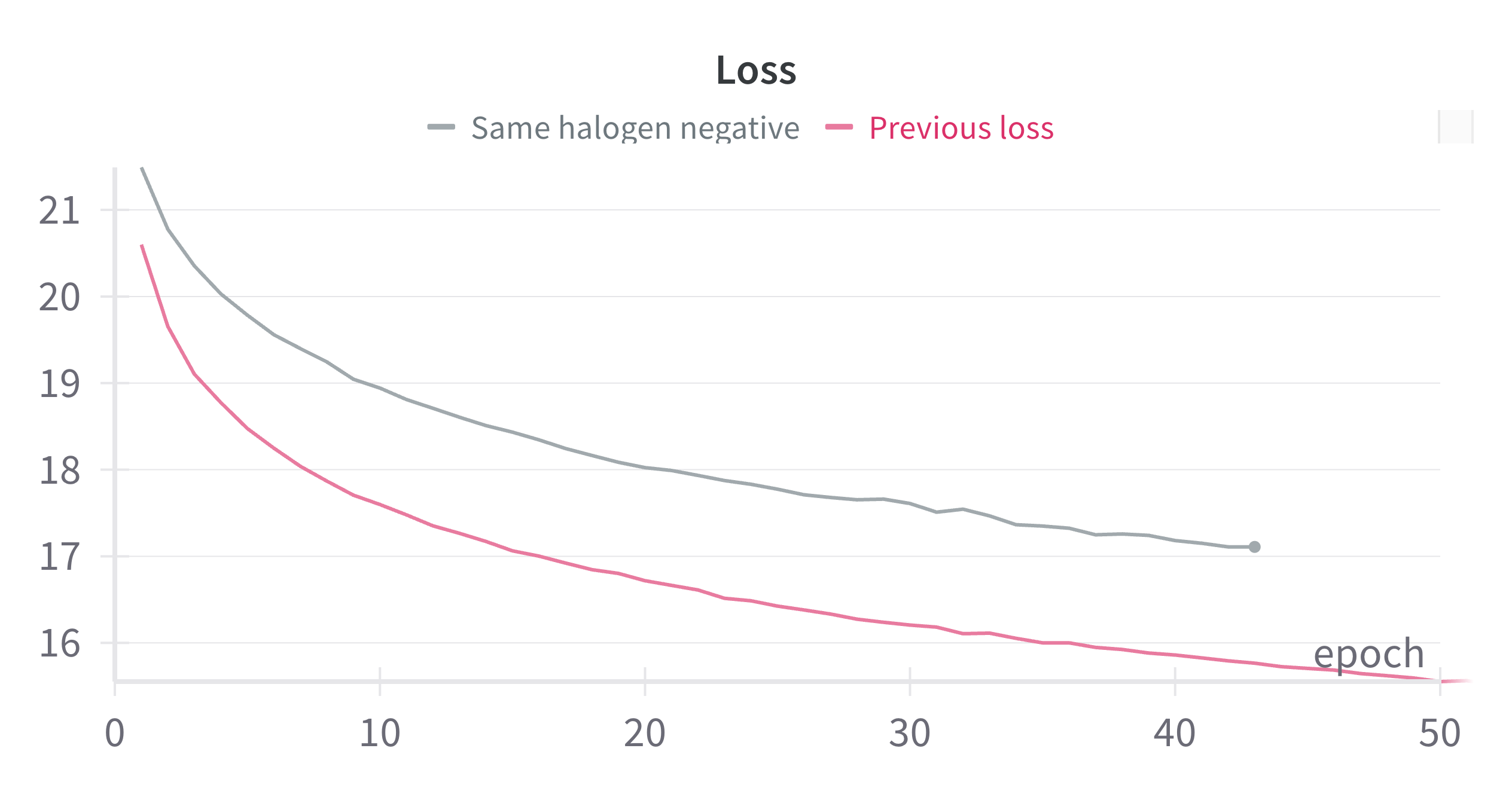}
        \caption{Total substrate scope contrastive loss.}
    \end{subfigure}
    \hfill
    \begin{subfigure}[t]{0.65\textwidth}
        \includegraphics[width=\textwidth]{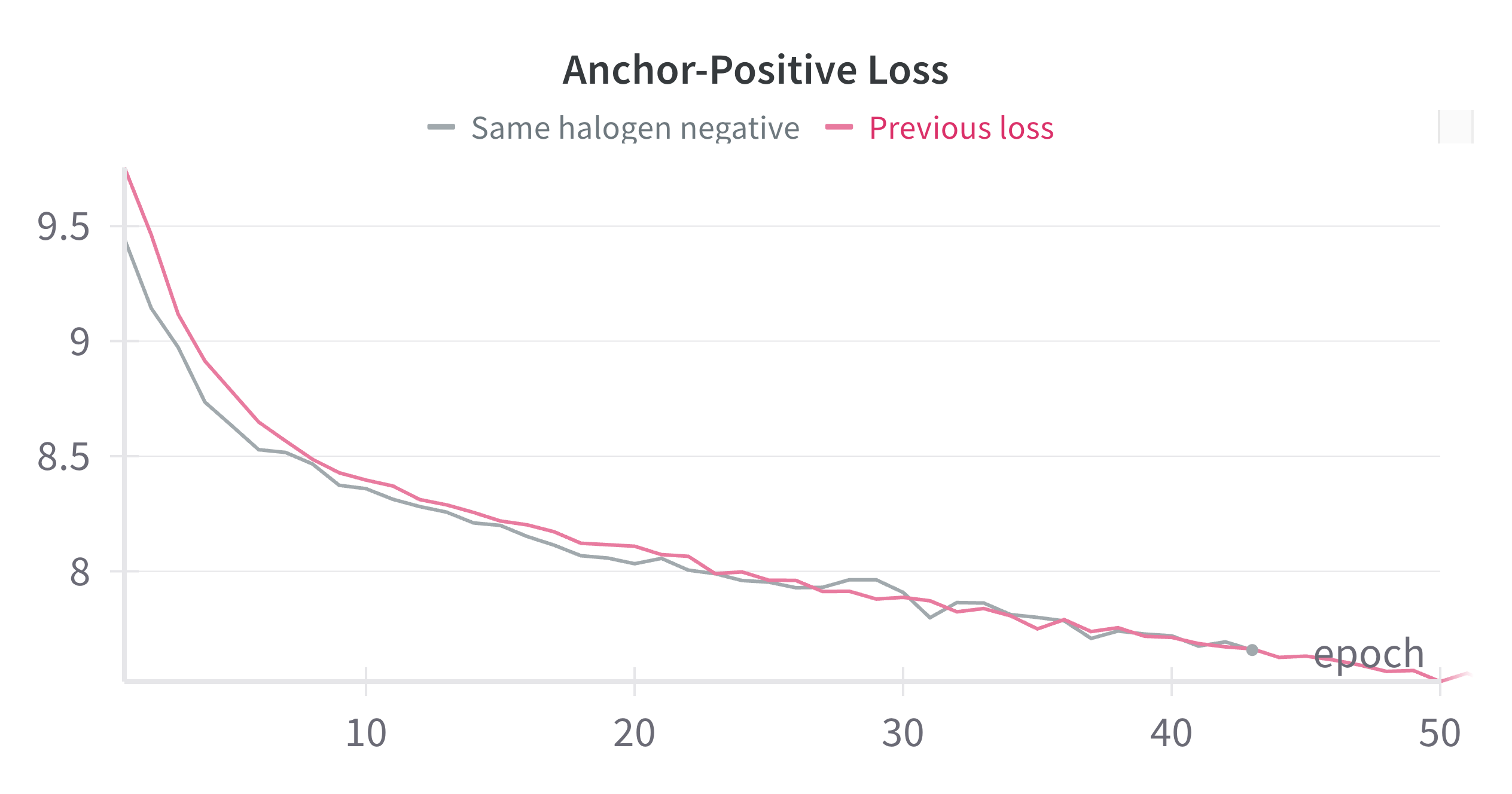}
        \caption{The value of the anchor-positive term.}
    \end{subfigure}
    \begin{subfigure}[t]{0.65\textwidth}
        \includegraphics[width=\textwidth]{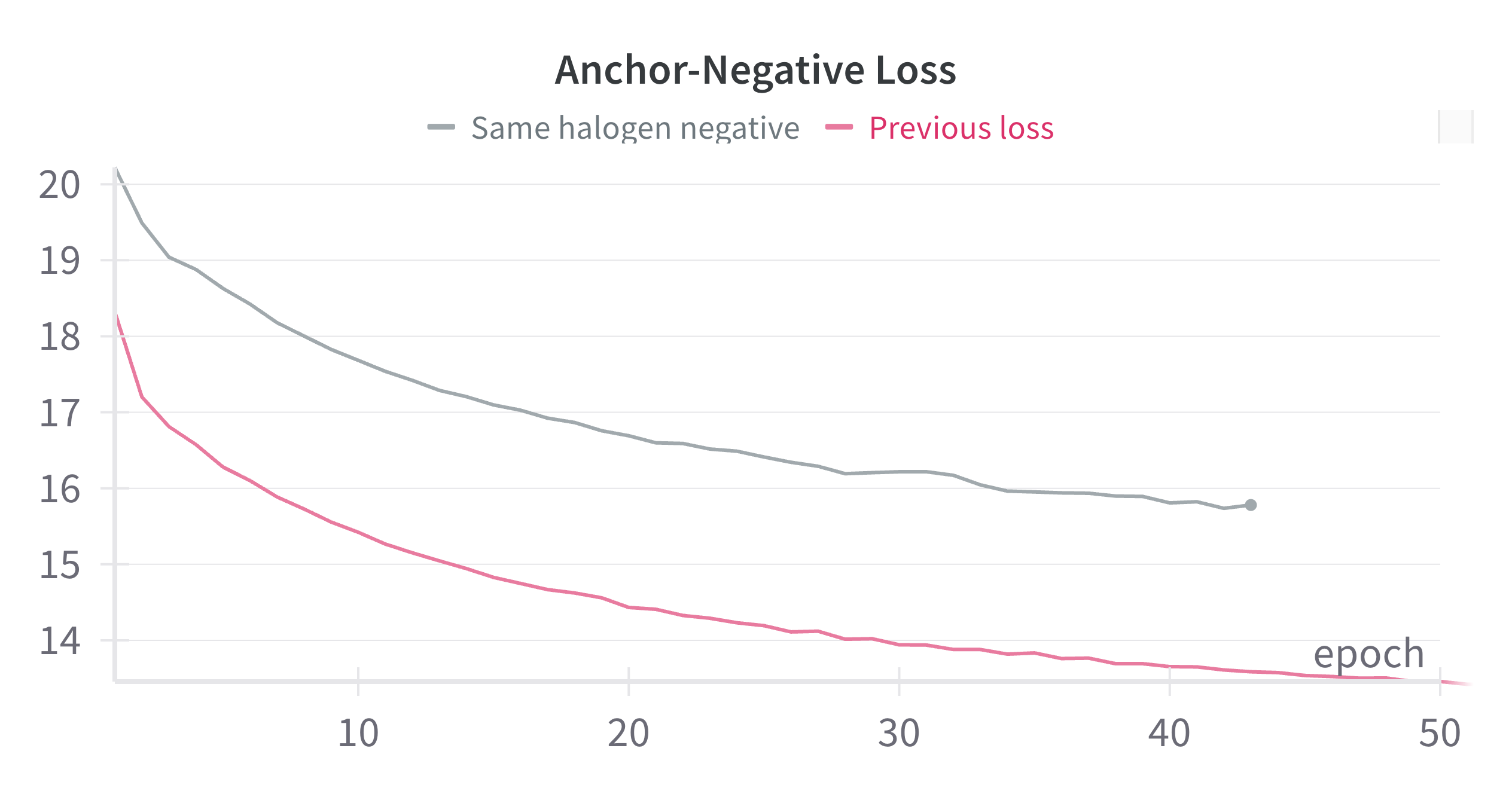}
        \caption{The value of the anchor-negative term.}
    \end{subfigure}
    \caption{The comparison of learning curves of the pre-training process with (Same halogen negative) and without (Previous loss) same halogen sampling.}
    \label{fig:shn}
\end{figure}

The impact of this strategy is illustrated in the learning curves shown in Figure \ref{fig:shn}. The results indicate that while the anchor-positive loss remains at a similar level and trend compared to the original sampling strategy, the anchor-negative loss is shifted upward. This shift suggests that the hard negative sampling strategy effectively introduces more challenging examples for the model to distinguish, validating its intended purpose. However, despite the effective negative sampling, we did not observe a quantitative improvement in performance metrics. We hypothesize that this arises from the limited modeling capacity of the specific network architecture employed in this study. This observation highlights the potential need for more expressive model architectures to fully leverage the benefits of hard negative sampling. We report the model trained with the original strategy in the main text and leave further exploration for future work.

\subsection{Substrate scope comparison}

\begin{figure}[H]
    \centering
    \includegraphics[width=\textwidth]{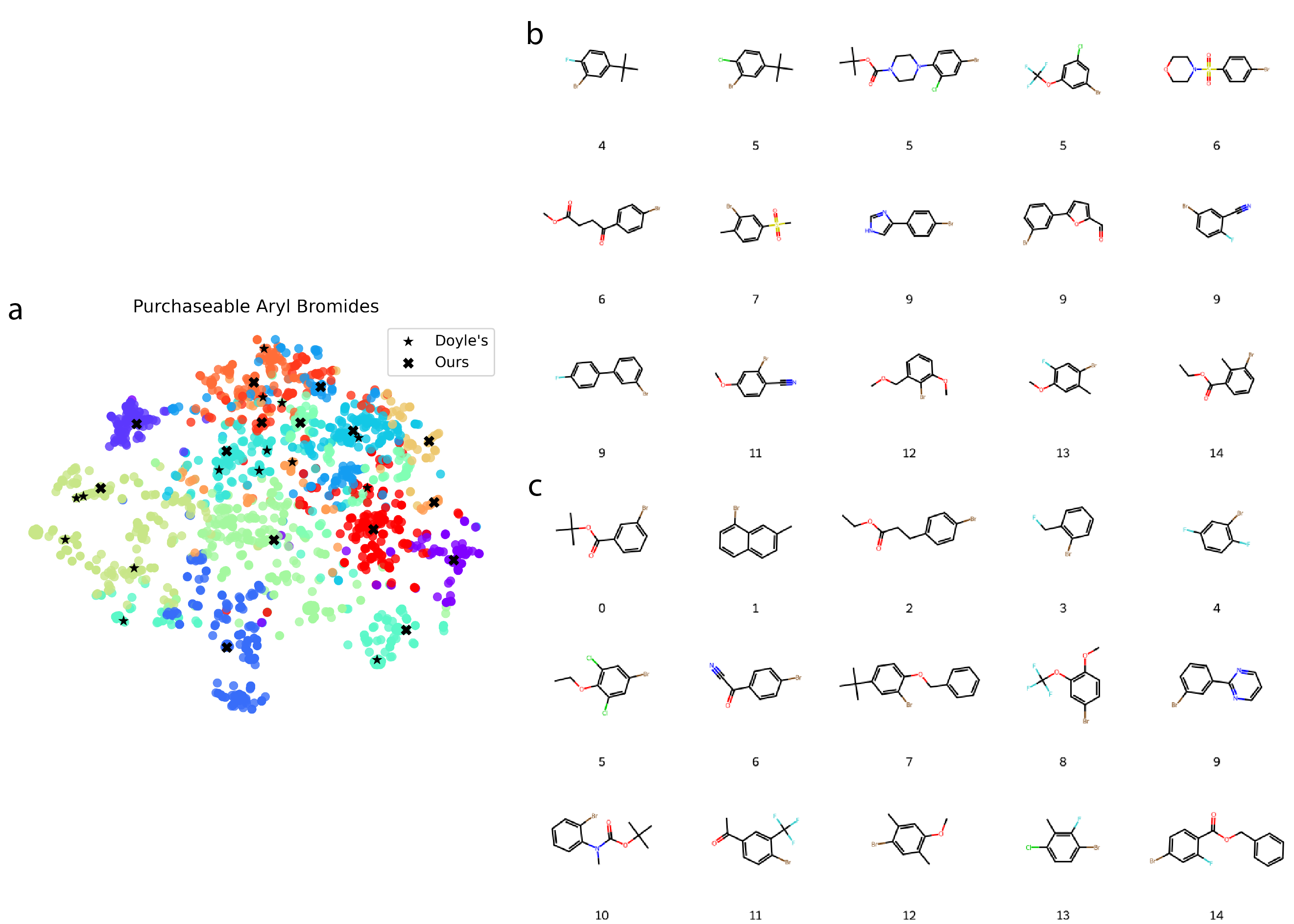}
    \caption{A comparison of substrate scope selected using DFT descriptors and our learned embeddings. (a) a t-SNE projection of all purchasable aryl bromides, categorized based on clustering. Within this projection, substrates selected via DFT descriptors are highlighted with star symbols (labeled as 'Doyle's'), whereas those chosen through our method are marked with cross symbols (denoted as 'Ours'). (b) chemical structures of substrates selected based on DFT descriptors, with accompanying numerical annotations for cluster identification. (c) chemical structures of substrates selected based on our learned embeddings, with accompanying numerical annotations for cluster identification.}
    \label{fig:Fig_scope_design}
\end{figure}

\textbf{Substrate scope selection.} We illustrate how learned embeddings can assist in the selection of a ``diverse'' set of aryl bromides, which one may wish to do at the outset of an experimental screening campaign. This process, inspired by \citeauthor{kutchukian2016chemistry}'s\cite{kutchukian2016chemistry} and \citeauthor{kariofillis2022using}'s\cite{kariofillis2022using} methodology, involved clustering the chemical space of commercially-available aryl bromides using K-means\cite{lloyd1982least}; while the authors used a DFT-derived feature vector, here we use our learned embeddings. A representative aryl bromide is chosen from each cluster to form a set designed to exhibit a diverse range of reactivities (Figure \ref{fig:Fig_scope_design}). 
Unlike clustering based on structural fingerprints or expert-selected descriptors, our approach relies on the specific reactivity profiles of aryl halides as exemplified by our embeddings, in principle leading to a more curated set for this class of compounds better aligned with their unique reactivity characteristics. As there is no objective correct answer for diverse substrate selection, we leave the selection for qualitative interpretation by the reader.

\end{document}